\documentclass[aps,10pt,prl,superscriptaddress,showpacs,floatfix,twocolumn,longbibliography]{revtex4-1}

\usepackage{graphicx}
\usepackage{epstopdf}
\usepackage{amsmath}
\usepackage{amssymb}
\usepackage{hyperref}
\usepackage{bm}
\usepackage{subfigure}

\hypersetup{
    colorlinks=true,       
    linkcolor=cyan,          
    citecolor=magenta,        
    filecolor=magenta,      
    urlcolor=cyan,           
    runcolor=cyan
}
\usepackage[pdftex]{color}

\newcommand{\beq}{\begin{equation}}
\newcommand{\eeq}{\end{equation}}
\newcommand{\beqnn}{\begin{equation*}}
\newcommand{\eeqnn}{\end{equation*}}
\newcommand{\bea}{\begin{eqnarray}}
\newcommand{\eea}{\end{eqnarray}}
\newcommand{\beann}{\begin{eqnarray*}}
\newcommand{\eeann}{\end{eqnarray*}}
\newcommand{\bes} {\begin{subequations}}
\newcommand{\ees} {\end{subequations}}

\newcommand{\ignore}[1]{}

\newcommand{\GS}{\mathrm{G}}
\newcommand{\ES}{\mathrm{E}}

\begin{document}
\title{Analog Errors in Ising Machines}
\author{Tameem Albash}
\affiliation{Information Sciences Institute, University of Southern California, Marina del Rey, California 90292, USA}
\affiliation{Department of Physics and Astronomy, University of Southern California, Los Angeles, California 90089, USA}
\affiliation{Center for Quantum Information Science \& Technology, University of Southern California, Los Angeles, California 90089, USA}

\author{Victor Martin-Mayor}
\affiliation{Departamento de F\'isica Te\'orica, Universidad Complutense, 28040 Madrid, Spain}
\affiliation{Instituto de Biocomputaci\'on y F\'isica de Sistemas Complejos (BIFI), Zaragoza, Spain}

\author{Itay Hen}
\affiliation{Information Sciences Institute, University of Southern California, Marina del Rey, California 90292, USA}
\affiliation{Department of Physics and Astronomy, University of Southern California, Los Angeles, California 90089, USA}
\affiliation{Center for Quantum Information Science \& Technology, University of Southern California, Los Angeles, California 90089, USA}
\email{itayhen@isi.edu}
\begin{abstract}
Recent technological breakthroughs have precipitated the availability of specialized devices that promise to solve NP-Hard problems faster than standard computers. These `Ising Machines' are however analog in nature and as such inevitably have implementation errors. We find that their success probability decays exponentially with problem size for a fixed error level, and we derive a sufficient scaling law for the error in order to maintain a fixed success probability. We corroborate our results with experiment and numerical simulations and discuss the practical implications of our findings.
\end{abstract}

\maketitle
%

{\bf Introduction}.---Recently there has been a flourishing of experimental quantum information processing devices~\cite{ibmNature,rigetti,Google9,Google49,DW2000}. The scientific community is readying itself for the first demonstration of `quantum supremacy'~\cite{preskill1,Bremner2017achievingsupremacy,PhysRevLett.118.040502,PhysRevLett.117.080501,PhysRevLett.112.130502,Broome794,Spring798,Google49} in the Noisy Intermediate-Scale Quantum (NISQ) era \cite{PreskillNISQ}, whereby quantum information processing devices will start performing tasks not accessible even for the largest classical supercomputers. At the forefront of this effort has been the development of analog quantum devices performing quantum annealing~\cite{kadowaki_quantum_1998,farhi_quantum_2000}, already realized on various  platforms such as arrays of superconducting flux qubits~\cite{LL1,LL2,LL3,DW2000,Johnson:2010ys,Berkley:2010zr,Harris:2010kx,Bunyk:2014hb}, to solve optimization problems, that is, to find bit assignments that minimize the cost of discrete combinatorial problems or equivalently the ground states (GSs) of Ising Hamiltonians. With this growing interest in developing special purpose devices for solving such problems, alternatives have also emerged~\cite{coherentIsing1,coherentIsing2,QNN}, touting improved performance over standard computers~\cite{2058-9565-2-4-044002,Inagaki603}.  

These Ising machines are analog in nature: the programmable parameters of the cost functions they aim to solve are controlled by continuous fields, implying that the intended cost function is only implemented to a certain precision. In this Letter, we study the effect of the analog nature of these devices on their ability to find global minima.  We argue that even when such devices are assumed ideal, i.e., always find a minimum of the \emph{implemented} cost function, the obtained configurations become exponentially unlikely to be global minima of the \emph{intended} problem.  In the absence of fault tolerant error correction, this represents a fundamental limitation to the scalability of Ising machines. We corroborate our analytical findings with numerical simulations as well as experiment carried out on a D-Wave 2000Q processor~\cite{King:2017}. We further derive a scaling law for how the magnitude of implementation errors must be reduced with problem size in order to maintain acceptable performance. We find that errors must scale as a power-law with problem size, with a model dependent exponent. Our results imply that fixed-size classical repetition codes are not a feasible approach to help these devices maintain their performance asymptotically with size.  
 
{\bf Analog Ising machines}.---Analog Ising machines suffer by nature from implementation errors caused by the conversion of the intended problem parameters to analog signals~\footnote{Other sources of errors not discussed here also exist such as $1/f$-noise~\cite{oneOverF2}.}. The Hamiltonian implemented by these devices is generally of the form:
\bea \label{eqt:Ising}
H_\sigma({s})&=& H_0 + \delta H_{\sigma} =  \sum_{\langle i j \rangle} \tilde{J}_{ij} s_i s_j +\sum_i \tilde{h}_i s_i \\\nonumber
&=&\sum_{\langle i j \rangle} \left(J_{ij} + \delta J_{ij} \right) s_i s_j +\sum_i \left(h_i  + \delta h_i \right) s_i   \,,
\eea
where $\{s_i=\pm 1\}$ are binary variables that are to be optimized over and $\langle i j\rangle$ denotes the underlying connectivity graph of the model. We denote by $\{J_{ij},h_i\}$ the {intended} Ising couplings between connected spins and the intended local fields on individual spins respectively of the intended Hamiltonian $H_0$, and we denote by $\{\tilde{J}_{ij},\tilde{h}_i\}$ the {implemented} values, obeying 
$\tilde{J}_{ij} = J_{ij} + \delta J_{ij}$ and $\tilde{h}_{i} = h_{i} + \delta h_i$.  The variables $\delta J$ and $\delta h$ represent the noise due to the analog nature of the device, and we take for simplicity $\delta J, \delta h \sim \mathcal{N}(0,\sigma^2)$~\cite{King:2015zr}.  We assume the noise for the different $\delta J_{ij}$ and $\delta h_i$ is statistically independent~\footnote{We expect our results to be robust to noise with a small local correlation but certainly not to noise with long-range statistical correlations.}.

To isolate the effect of analog noise on performance \cite{venturelli2015,martin-mayor:15,oneOverF1,PhysRevA.94.022337}, where success is defined as finding a GS of the intended Hamiltonian, we assume that the machine is otherwise {ideal}, i.e., that it {always} finds the GS of the implemented Hamiltonian. This allows us to draw from classical spin glass theory, where it is known that spin glasses are susceptible to a phenomenon referred to as $J$-chaos, in which perturbations to the intended Hamiltonian change the identity of the GS~\cite{nifle:92,ney-nifle:98,krzakala:05,katzgraber:07}.

{\bf Analytical treatment}.---
For concreteness, we consider a spin-glass Hamiltonian of the form of Eq.~\eqref{eqt:Ising}, where the intended couplings take values $J_{ij}= \left\{1,0,-1 \right\}$ and $h_i =0$~\footnote{Our derivation can be straightforwardly generalized to other models as  well.}. 
Let us denote a GS of the intended Hamiltonian (in general, there could be exponentially many of those) by ${s}^{\GS}$ and an excited state (ES) of the intended Hamiltonian by ${s}^{\ES}$.   The energy gap between the two is
\beq
\Delta E_0 = H_0({s}^{\ES}) - H_0({s}^{\GS}) > 0 \ .
\eeq
The ES has a chance of becoming the GS of the implemented Hamiltonian only if
$
\Delta E_{\sigma}=H_{\sigma}({s}^\ES)- H_{\sigma}({s}^\GS)<0. 
$
Expressed differently, 
\begin{eqnarray}
\Delta E_{\sigma} &=& \Delta E_{0} - \sum_{\langle i j \rangle}\delta J_{ij} s_{i}^{\GS}s_{j}^{\GS} (1-q^{\mathrm{link}}_{ij}) \nonumber \\
&&- \sum_i \delta h_i s_i^\GS ( 1 - q_i ) \,,\label{eq:Hdiff-qlink}
\end{eqnarray}
where we have introduced the spin overlap $q_{i} = s_i^\GS s_i^\ES$ and the link-overlap $q_{ij}^{\mathrm{link}}=s_i^{\GS} s_j^{\GS}\,s_i^{\ES}s_j^{\ES}$ corresponding to the bond $(ij)$.
When $q^{\mathrm{link}}_{ij}=1$, it implies that if the bond $(ij)$ is satisfied (unsatisfied) by the GS of the intended Hamiltonian, it is also satisfied (unsatisfied) by the ES. Similarly $q^{\mathrm{link}}_{i,j}=-1$ implies that the corresponding bond is satisfied in one spin assignment and unsatisfied for the other. 

We now define $W$ to be the total number of bonds with $q^{\mathrm{link}}_{ij}=-1$
and $D$ to be the total number of spins with $q_i = -1$ (or equivalently the Hamming distance between the ES and the GS).
Because (i) the $\delta J_{ij}$ and $\delta h_{i}$ are statistically independent from $s_i^{\GS}$ and $s_i^{\GS}s_j^{\GS}$ and (ii) the $\delta J_{ij}$ for different bonds and $\delta h_i$ for different spins are mutually independent, we can write
\beq
\Delta E_{\sigma}=\Delta E_0+2\sigma \sqrt{W + D}\, \eta \,,\label{eq:crucial-eta}
\eeq
where $\eta$ is a normal random variable $\mathcal{N}(0,1)$. Defining the parameter $z=\frac{\Delta E_{0}}{2\sigma\sqrt{W + D}}$, the probability $p(z)$ of having $\Delta E_{\sigma}<0$ is given by
\begin{eqnarray}
p(z) \equiv \mathrm{Prob}[\Delta E_\sigma < 0] &=&\int_{-\infty}^{-z} \frac{\mathrm{d}\eta}{\sqrt{2\pi}}\,\mathrm{e}^{-\eta^2/2}\, \ . \label{eq:pz-def}
\end{eqnarray}
Therefore, the likelihood that a particular ES lies below a GS for
the implemented Hamiltonian ranges from being
infinitesimal [more precisely $p(z)\sim \mathrm{e}^{-z^2/2}$ for $z\gg 1$], to
50\% (for $z\ll 1$). In fact, $p(z)$ become
sizable at $z=1$.

For physical devices, the number of couplers scales linearly with number of spins $n$, so at most $W\sim n$. Similarly, $D$ can at most scale as $\sim n$.
More specifically, we identify two kinds of ESs (see Ref.~\cite{martin-mayor:15}). On the one hand, there are `topologically trivial' ESs for which $W\sim 1$. For these, there are no energy barriers separating them from a GS, i.e., gradient descent takes the ES to a GS and therefore, $p(z)$ is extremely small for these ESs up to $\sigma\sim 1$. On the other hand, there exist topologically non-trivial ESs for which $W\sim n$. 
These low lying ESs produce a sizable $p(z)$ already for $\sigma\sim 1/\sqrt{W + D}\sim 1/\sqrt{n}$. From now on, we consider only the latter kind.

So far we have only considered a pair of states, a single GS and a single ES (both out of possibly exponentially many \cite{barahona:82,PhysRevLett.96.167205,Tchaos1}). However, the relative degeneracies of these must be taken into account. To do that, in lieu of relying on a particular model~\cite{krzakala:05}, we make a few simplifying assumptions.
First, let us restrict to the ESs with a fixed
$\Delta E_0$ (in our case  $\Delta E_0 \geq 2$)  and assume there are $N_\mathrm{ES}$ of those.
 We further assume that:
i) $W$ does not fluctuate significantly between different ESs. In other words,
  $p(z)$ defined in Eq.~\eqref{eq:pz-def} can be roughly regarded as
  non-fluctuating.
ii)  In Eq.~\eqref{eq:crucial-eta}, each ES defines a random variable $\eta$. We now assume that the $\eta$'s for different ESs can be regarded as independent. In other words, $N_\mathrm{ES}$ becomes the \emph{effective}
number of statistically independent $\eta$'s within the entire
population of ESs of the intended Hamiltonian (we later show that the conclusions that follow from these assumptions are consistent with numerical results).

Under these assumptions, the probability of a single GS to remain energetically favorable to \emph{every} ES  for
the implemented Hamiltonian is
\beq\label{eq:p-1GS}
p_{1\,\mathrm{GS}}\equiv [1-p(z)]^{N_{\mathrm{ES}}}\, \ .
\eeq
Since we expect $N_{\mathrm{ES}}$ to scale at least polynomially (if not exponentially) with system size, we conclude that the likelihood for a single GS to remain optimal for the implemented Hamiltonian \emph{decays} exponentially with system size.
 
In the most general case, there may be an \emph{effective} number of GSs, $N_\mathrm{GS}$, only one of which is required to remain optimal for an analog Ising machine to succeed.  
The probability that at least one GS remains more optimal than all the ESs for the implemented Hamiltonian is:
\beq \label{eqt:Success}
p_{\mathrm{S}} = 1 - [1-p_{1\,\mathrm{GS}}]^{N_{\mathrm{GS}}} \ .
\eeq
We expect that the growth of $N_{\mathrm{GS}}$ is unlikely to overcome the decay of $p_{1\,\mathrm{GS}}$ with system size~\footnote{We can always make $p_{1\,\mathrm{GS}}$ smaller than $1/N_{\mathrm{GS}}$ by increasing $\sigma$ for example.  This will increase the typical $\Delta E_0$, which implies that the typical $N_{\mathrm{ES}}$ will grow.}, and it would then follow that $p_\mathrm{S}$ will also become exponentially small with system size, i.e., analog Ising machines are expected to fail for any fixed noise level as we scale up the system size.

At this point, we address how must the noise strength $\sigma$ scale in order to still have a sub-exponential probability to succeed.  In the regime where the asymptotic estimate $p(z)\sim\mathrm{e}^{-z^2/2} \ll 1$ holds, i.e. a regime where the failure rate is small, rewriting Eq.~\eqref{eqt:Success} for a fixed $p_{\mathrm{S}}$ requires that $z$ behave as:
\beq
z^2 \approx 2 \left( \log(N_{\mathrm{ES}}) - \log(1 - p_{\mathrm{S}}) /N_{\mathrm{GS}} \right) \ .
\eeq
If in addition the effective number of ESs scales as $N_{\mathrm{ES}} \sim \mathrm{e}^{B n^k}$ (with $k = 0$ denoting sub-exponential scaling) \footnote{While studies on temperature chaos~\cite{Tchaos1} suggest that ESs are exponentially more abundant than GSs~\cite{martin-mayor:15} which might suggest that $k=1$, here we are requesting topologically non-trivial and statistically independent ESs.}, 
then we conclude the noise level must be scaled down as
\begin{equation} \label{eqt:scaling}
\sigma_{\mathrm{S}} \sim 1/\sqrt{(W + D) n^k} \ .
\end{equation}
A worst-case estimate would be to take $W,D \sim n$ and $k=1$, in which case an estimate of the required scaling of the noise to maintain a sub-exponential success probability is
\begin{equation} \label{eqt:worsecase}
\sigma_{\mathrm{S}}^{\mathrm{(worst \ case)}} \sim 1/n\, .
\end{equation}
{\bf Numerical and experimental results}.---
To illustrate the validity of the above derivation and accompanying simplifying assumptions, we present results of numerical simulations done on instances with known GSs.  Our first problem class is of planted-solution instances~\cite{Hen:2015rt} defined on a Chimera lattice: an $L \times L$ grid of 8-spin unit cells in a $K_{4,4}$ formation and is the graph of currently available D-Wave quantum annealers~\cite{Choi1}. 
The instances have a degenerate GS corresponding to all spins pointing up or all spins pointing down (due to the $Z_2$ symmetry of the problem Hamiltonian).  The couplings are restricted to have magnitudes $\{1/3, 2/3, 1\}$.  (See the SM for further details~\footnote{In the SM, we consider two additional problem classes, namely, 3-regular 3-XORSAT instances involving 3-body couplings, which exhibits a similar behavior, and Ising instances defined on a square grid with random $\pm 1$ couplings}.)  

To simulate analog errors, we introduce a zero-mean Gaussian noise with standard deviation $\sigma$ as in Eq.~\eqref{eqt:Ising} to  the implemented Hamiltonian, and we use the Hamze-Freitas-Selby (HFS)~\cite{hamze:04,Selby:2014tx} algorithm to find the GSs of the implemented Hamiltonian. While the HFS algorithm is not guaranteed to find the GS with certainty, we can know with certainty if the GS of the implemented Hamiltonian has changed from that of the desired Hamiltonian if (1) the returned state by the algorithm has a lower energy on the implemented Hamiltonian than the two GSs of the intended Hamiltonian, and (2) the returned state is an ES of the intended Hamiltonian. In order to ensure that we only consider topologically non-trivial ESs, we perform 100 sweeps of steepest-gradient updates: we flip each spin, and if the energy is reduced (according to the intended Hamiltonian) we accept the update and reject otherwise.  In cases where the returned state fails either condition, we assume that the GS of the intended Hamiltonian has not changed.  Since this may in fact be incorrect, our results are an upper bound on the probability that the GS has not changed.  For each grid size $L \in [8,14]$, we generate $10^2$ instances representing our intended Hamiltonians, and for each instance we generate $10^3$ noisy realizations of the implemented Hamiltonian.

\begin{figure}[t] 
   \centering
      \includegraphics[width=.87\columnwidth]{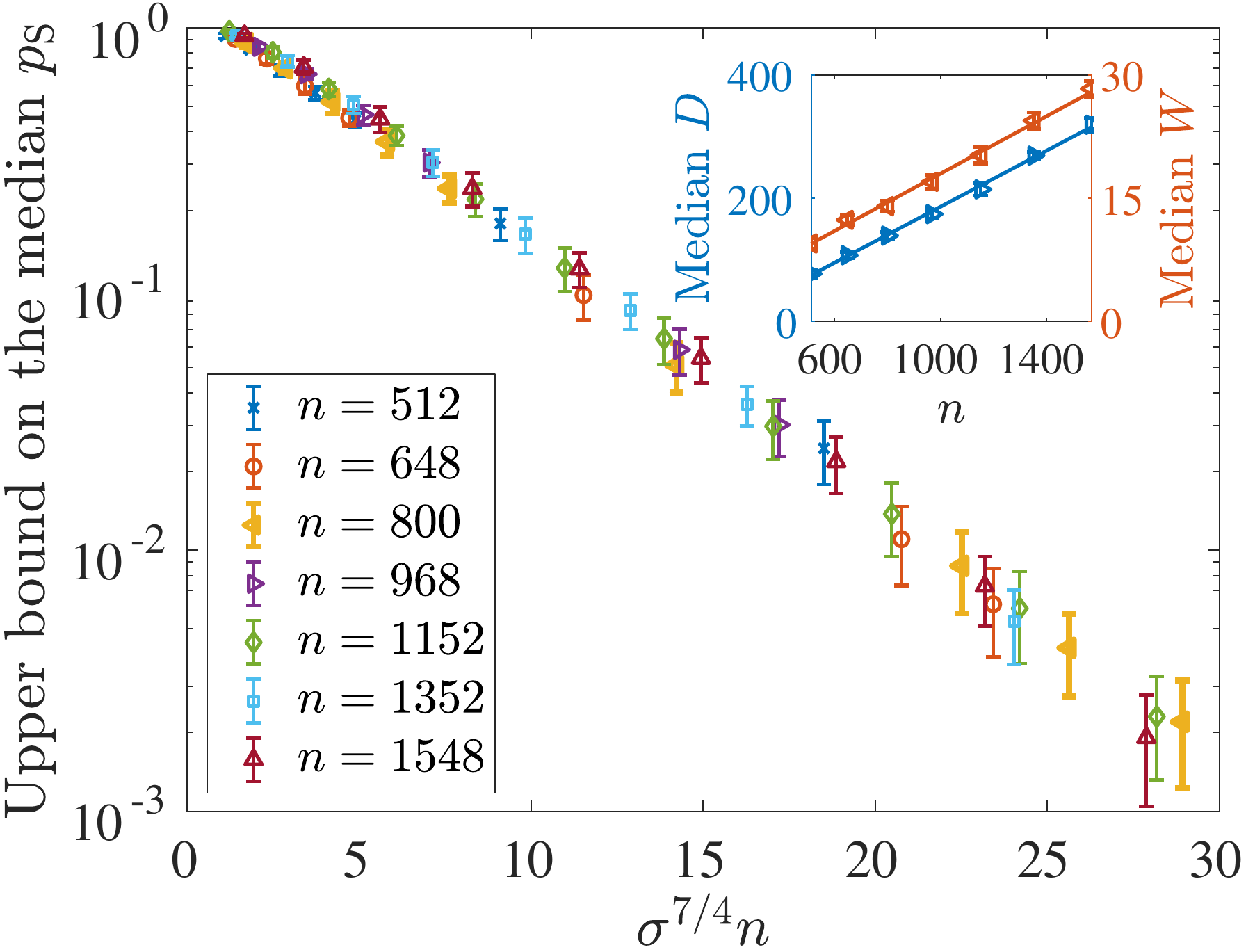} 
   \caption{Upper bound on the median probability $p_{\mathrm{S}}$ over $10^2$ instances that the GS of the intended Hamiltonian remains the GS of the implemented Hamiltonian for varying planted-solution instances on different problem sizes $n$ and Gaussian noise strengths $\sigma \in \left[ 0.01,0.2\right]$.  Error bars correspond to two standard deviate error bars obtained from bootstrapping over the $10^2$ instances. Inset: the scaling of the median $D$ and $W$ from the intended GSs for $\sigma = 0.1$.  Solid line corresponds to a linear fit with slope $\alpha \approx 0.23$ and $\alpha \approx 0.02$ for $D$ and $W$ respectively.}
   \label{fig:ChimeraCollapse}
\end{figure}

In Fig.~\ref{fig:ChimeraCollapse}, we show the probability that the GS of the intended Hamiltonian remains the GS of the implemented Hamiltonian for several sizes and noise strengths. Our results show an exponential dependence as a function of $\sigma^2 n$ for large $p_\mathrm{S}$ and approaching $\sigma^{7/4} n$ for small $p_{\mathrm{S}}$, consistently with the above analytical derivation.  This change in scaling may be expected since as the system size or noise level grows the effective number of ESs grows as well.
 We find that for these instances $W$ and $ D$ scale linearly with $n$ (shown in the inset of Fig.~\ref{fig:ChimeraCollapse}), which suggests that the number of effective ESs grows from polynomial to a stretched-exponential with $k=1/7$ [Eq.~\eqref{eqt:scaling}].  For these instances then, the noise level must scale as $\sigma_{\mathrm{S}} \sim n^{-4/7}$ in order to maintain a fixed performance level.

\begin{figure}[t]
\begin{center}
\includegraphics[width=.87\columnwidth]{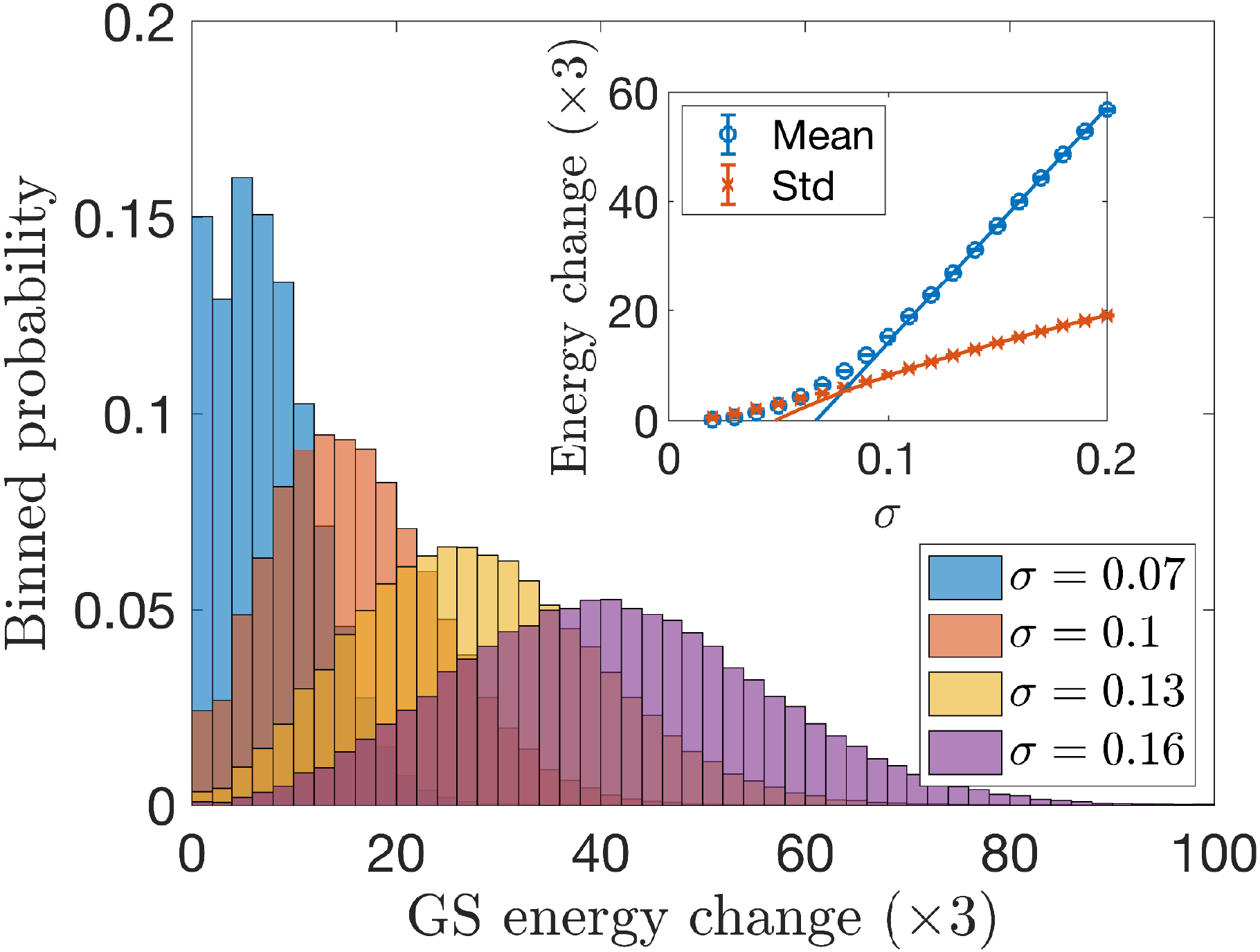}
\caption{{Histogram of the energy of the GSs of the implemented (noisy) Hamiltonian as measured by the intended Hamiltonian.}  Here we use $10^3$ noise realizations for each of the $10^2$ planted-solution Chimera instances defined on a $12\times 12$ grid.  Only the topologically non-trivial ESs of the intended Hamiltonian are shown.  Inset: the scaling of the mean and standard deviation of the distribution of GS energies.  The error bars correspond to two standard deviate error bars generated by a bootstrap over the $10^5$ data points.  The solid curves are fits to $\mu_E = a + \sigma b$ and $\sigma_E = c + \sqrt{\sigma} d$ for the mean and standard deviation respectively, with $a = -28.82, b = 429.47$ and $c = -18.57, d = 84.28$.}
\label{fig:enDist}
\end{center}
\end{figure}

While our analysis so far has focused only on whether the intended GS is no longer a GS of the implemented Hamiltonian, it is instructive to study the energy distribution (according to the intended Hamiltonian) of the topologically non-trivial ESs that become the new GSs of the implemented Hamiltonian.  This gives us a sense of how far away in energy the implemented GSs are from the intended GS. We show this in Fig.~\ref{fig:enDist}, where we see that for a fixed system size and growing noise strength, the distribution becomes more Gaussian-like with the mean moving farther away from the GS energy.  Similar behavior occurs for a fixed noise strength and growing system size, which we provide in the SM.  This effect is similar to the behavior of a thermal energy distribution for a fixed temperature and increasing system size~\cite{PhysRevLett.119.110502}.

\begin{figure}[htbp] 
   \centering
   \includegraphics[width=.87\columnwidth]{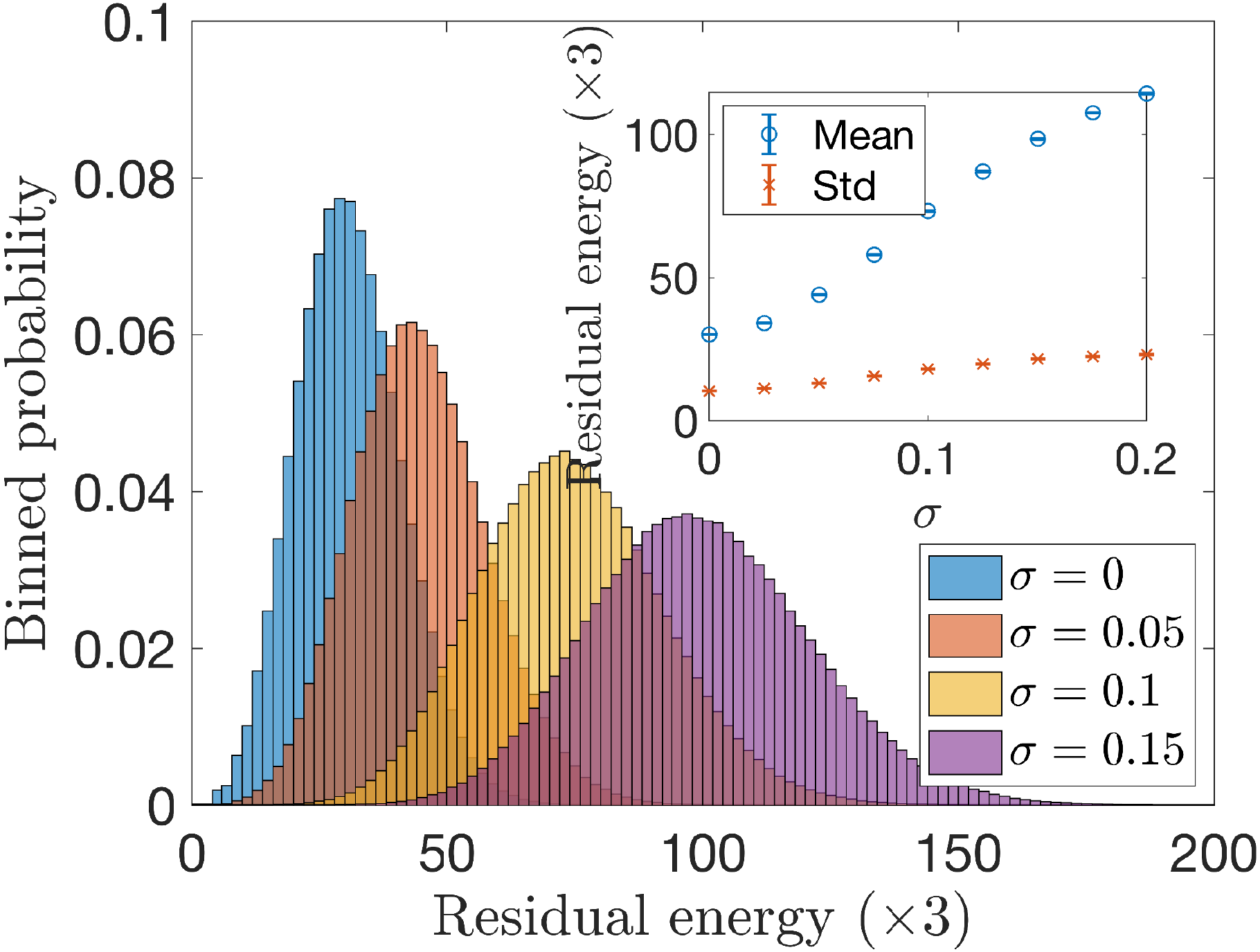} 
   \caption{Residual energy of the states returned by the D-Wave 2000Q quantum annealing processor at NASA's Ames Research Center after post-processing with 100 sweeps of steepest descent for a single instance.  The instance here is defined on a $12\times12$ subgraph of the processor (a total possible of $n = 1142$ spins), and it is generated in an identical way to the Chimera instances studied in Figs.~\ref{fig:ChimeraCollapse} and \ref{fig:enDist}.  Gaussian noise with mean 0 and standard deviation $\sigma$ is added to both the couplings and local fields before the instance is submitted to the processor.  For this single instance, $10^4$ noise realizations were generated and $10^2$ anneals were performed for each noise realization.  (Further details are given in the SM.)}
   \label{fig:DW2KQ}
\end{figure}

We corroborate our analysis with runs on a D-Wave 2000Q quantum annealing processor \cite{DW2000}.  The device has an intrinsic noise level~\cite{King:2014uq,martin-mayor:15}, however we introduce additional Gaussian noise to simulate the effect of different noise levels.  As shown in Fig.~\ref{fig:DW2KQ}, even in the absence of noise, the device rarely finds a GS, which we can attribute to both thermal \cite{PhysRevLett.119.110502} and analog errors on the device.  As the noise level is increased, the distribution of states observed moves even farther away from the GS energy, in a manner consistent with Fig.~\ref{fig:enDist}.  

{\bf Conclusions and outlook}.---
In this work, we have pointed to a fundamental limitation of analog Ising Machines, whereby implementation errors detrimentally affects their scaling performance.
 We have shown that even under the assumption that these devices are otherwise ideal, i.e., that they instantaneously find a minimizing configuration of the implemented problem, their success probability decays exponentially with system size for any fixed nonzero noise level.  Such errors have important ramifications beyond the context of optimization; in the setting of generating thermal (Boltzmann) samples, we can expect that analog errors may distort the sampled distribution \cite{Hauke:2012}.  Even quantum logic gates cannot be implemented perfectly, but the difference is that fault-tolerant error correction can correct for these errors \cite{Knill:1997kx}.
 
We emphasize that our results do not mean that `disorder-chaos' should be expected for every mildly-disordered system. (We provide an example in terms of the 1d chain in the SM that is meant to illustrate this.) The key point is the nature of the low-energy landscape of the problem under consideration. Broadly speaking, `disorder-chaos' requires the problem Hamiltonian to have a \emph{large} number of very low-energy excitations, each of them wildly differing from the GS (i.e. Hamming distance of order $n$). Under such circumstances, it is intuitively clear that even tiny errors may cause one of these non-trivial excited states to take over as the true GS of the implemented Hamiltonian. In fact, the hypothesis we made to derive Eqs.~(\ref{eq:Hdiff-qlink}-\ref{eqt:Success}) made quantitative the above-outlined physical picture. 
 
 Now, it turns out that disorder chaos is a generic feature of spin-glasses. Because spin-glasses are an archetype of \emph{hard} optimization problems, we should expect to encounter disorder chaos in every optimization problem hard-enough to deserve to be solved with a specialized analog device such as an Ising machine. Indeed, the 3-regular 3-XORSAT instances that we study in the SM exhibit disorder chaos as well. Fortunately, as universality suggests, the scaling laws that we find for 3-XORSAT seem compatible with our findings for instances on the Chimera lattice. 
 
 Our results for the Chimera instances differ slightly from the results on $J$-chaos (sometimes also called bond-chaos) for ground state of the 2d Edward-Anderson model by Krzakala \& Bouchaud \cite{krzakala:05} (a finite temperature analysis was carried out in Ref.~\cite{PhysRevB.93.224414}), where they found a scaling of $\sigma^2 n$ for the regime of fully developed $J$-chaos.  This scaling suggests a sub-exponential scaling for the number of ESs [$k=0$ in Eq.~\eqref{eqt:scaling}].   This discrepancy might be accounted for by the fact, at exactly zero temperature, that the ruling fixed point for the Renormalization Group flow for Gaussian and binary couplings is different \cite{amoruso:03}.  Extending beyond 2d, the example of 3-regular XORSAT, which we present in the SM, exhibits a scaling closer to $\sigma^{4/3} n$, indicating a value of $k = 1/2$ for this class of instances. By deriving appropriate scaling laws, we have found that in order to counteract this reduction in performance as the system size grows, the noise level must be scaled down as a power law (with the worst case being $1/n$).  

We can consider how known error correction schemes would effectively reduce the magnitude of this noise. One way is to use classical repetition codes that implement multiple copies of the intended Hamiltonian~\cite{Young:2013fk,PAL:13,PAL:14,Vinci:2015jt,NQAC:2016}.  This has the feature of effectively rescaling the implemented Hamiltonian norm by a factor $K$ and hence effectively reducing the noise strength by a factor of $1/\sqrt{K}$~\cite{Young:2013fk}.  Therefore, a rescaling of $K\sim n^2$ is needed to achieve the necessary noise reduction for the worst case [Eq.~\eqref{eqt:worsecase}] and of $K\sim n^{8/7}$ for the Chimera example studied numerically here (Fig.~\ref{fig:ChimeraCollapse}).  More generally, for a $K$ that scales as $n^\alpha$, the encoded Hamiltonian using classical repetition schemes~\cite{NQAC:2016,Young:2013fk} would require a Hamiltonian for which the energy scale grows faster than linear with $n$.  For scalable architectures, this would require the number of physical spins per encoded spin to grow as a power law with $n$.  This highlights the importance of scalable error correction to maintain the performance of an algorithm as the system size scales.

We conclude by emphasizing that our analysis is asymptotic in nature and is valid in the large $n$ limit.  If only a specific problem class of a certain finite size is of interest, then it would be possible in principle to engineer a device with a sufficiently low noise for that problem class and size. It remains to be seen whether analog Ising machines will be a viable alternative to digital Ising simulations in light of the fundamental problems illustrated by our work. 
\\

\begin{acknowledgments}
{\bf Acknowledgements}.---
Computation for the work described in this paper was supported by the University of Southern California's Center for High-Performance Computing (hpc.usc.edu) and by the Oak Ridge Leadership Computing Facility, which is a DOE Office of Science User Facility supported under Contract DE-AC05-00OR22725. The research is based upon work partially supported by the Office of
the Director of National Intelligence (ODNI), Intelligence Advanced
Research Projects Activity (IARPA), via the U.S. Army Research Office
contract W911NF-17-C-0050. The views and conclusions contained herein are
those of the authors and should not be interpreted as necessarily
representing the official policies or endorsements, either expressed or
implied, of the ODNI, IARPA, or the U.S. Government. The U.S. Government
is authorized to reproduce and distribute reprints for Governmental
purposes notwithstanding any copyright annotation thereon. We acknowledge the support of the Universities Space Research Association (USRA) Quantum AI Lab Research Opportunity Program. V. M.-M. was partially supported by MINECO (Spain) through Grant No. FIS2015-65078- C2-1-P (this contract partially funded by FEDER).
\end{acknowledgments}
%
%

\onecolumngrid
\newpage
\begin{center}
\textbf{\large{Supplemental Information for}}\\
\textbf{\large{``Analog Errors in Ising Machines"}}\\
\end{center}

\twocolumngrid

\section{Chimera Instances} \label{app:Chimera}
In this work we have chosen one problem class to be that of the `planted solution' type---an idea borrowed from constraint satisfaction (SAT) problems. In this problem class, the planted solution represents a ground-state configuration of the Hamiltonian that minimizes the energy and is known in advance. The Hamiltonian of a planted-solution spin glass is a sum of terms, each of which consists of a small number of connected spins, namely, $H=\sum_j H_j$\cite{Hen:2015rt}.  Unlike previous work using planted-solution instances \cite{Hen:2015rt,King:2015zr,King:2017}, here we generate each instance such that every loop has only two ground states, the all-spin-up and all-spin-down configurations.  This is done by picking each term $H_j$ to be two randomly generated loops that share a single edge.  The common edge to the two loops is taken to be antiferromagnetic with magnitude 1, while the remaining edges of the two loops are taken to be ferromagnetic with magnitude 1.  We enforce that each loop of the pair must leave the Chimera cell and have length 6, and they are only added to the Hamiltonian $H$ if there additional does not result in a coupler with magnitude greater than 3 \cite{King:2015zr}.  We use a total of $\lceil 0.13 \times 8L^2 \rceil$ of these paired-loops for each instance.  The couplings $J_{ij}$ are finally rescaled so that the largest magnitude coupler has value 1. Under this rescaling, the smallest magnitude coupler will be 1/3.  We summarize properties of these instances in Fig.~\ref{fig:UniqueGSv6b}.  Critically, over the range of sizes we study, the typical fraction of couplers with magnitude 1/3, 2/3, and 1 remains  constant.

In order to determine the scaling behavior of the median $p_{\mathrm{S}}$ in Fig.~1 of the main text, we perform the following procedure.  We fix a probability $p$ and determine for each problem size $n$ the ${\sigma}_p(n)$ value for which $p_{\mathrm{S}}({\sigma}_p(n),n) = p$.  The scaling of ${\sigma}_p(n)$ with $n$ as we reduce $p$ then gives us the asymptotic scaling behavior we expect.  We show examples of this fitting procedure in Fig.~\ref{fig:FittingToDataChimera}.

We reproduce the reported linear behavior with $n$ of $W$ and $D$ for the topologically non-trivial ESs scale from the main text in Fig.~\ref{fig:Chimera_W_D}.  We also show in the same figure how the energy (according to the intended Hamiltonian) of these ESs scales linearly with $n$.  To complement these figures, we also show how these quantities behave with $\sigma$ for a fixed $n$ in Fig.~\ref{fig:Chimera_W_D2}.  We see that $W$ and $\Delta E_0$ have a dependence that goes as $\sim \sigma^2$ for small $\sigma$ but transitions to linear for large $\sigma$, although we may expect this behavior to change at very large $\sigma$ values since the mass $W$ is bounded by the number of couplers and similarly for the the energy distribution.  The behavior of the Hamming distance $D$ approaches the median value of the number of spins divided by 2 for large $\sigma$ (see Fig.~\ref{fig:ChimeraSpins} for the median number of spins for these instances).  

While our analytical treatment in the main text restricted to ESs with $\Delta E_0 \sim 1$ for concreteness, the scaling of $\Delta E_0$ with $n$ of the ESs that become the GS of the implemented Hamiltonian observed is not in contradiction with our analysis.  The analysis we present in the main text is not about which ES becomes the ground state, but simply about when $\Delta E_{\sigma}$ becomes $< 0$ which we claim occurs when $z \sim 1$.  We corroborate this by showing the behavior of the $z$ value of the topologically non-trivial ESs that become the GS of the implemented Hamiltonian in Fig.~\ref{fig:Chimera_z}.  We see that over the entire range of sizes $n$ and noise levels $\sigma$ studied we have $z \gtrsim 1$.

In the main text, we showed that for a fixed problem size, the states that become the ground state of the implemented Hamiltonian are at larger and larger energies as the noise scale is increased (Fig.~2 of the main text).  Here we show in Fig.~\ref{fig:enDist2} that a similar effect happens if we hold the noise level fixed and increase the problem size $n$.  We observe that as we increase $n$, the distribution of energies (as measured by the intended Hamiltonian) of the GS of the implemented Hamiltonian move further away from the GS energy of the intended Hamiltonian and become more Gaussian-like.  We find that the mean of this distribution scales linearly with $n$ and the standard deviation scales as $\sqrt{n}$.   This is consistent with the result shown in Fig.~\ref{fig:ChimeraMedianEnergy}.
%

\section{Chimera Instances on the D-Wave 2000Q Processor}
\begin{figure*}[htbp] 
   \centering
   \includegraphics[angle=270,width=1.5\columnwidth]{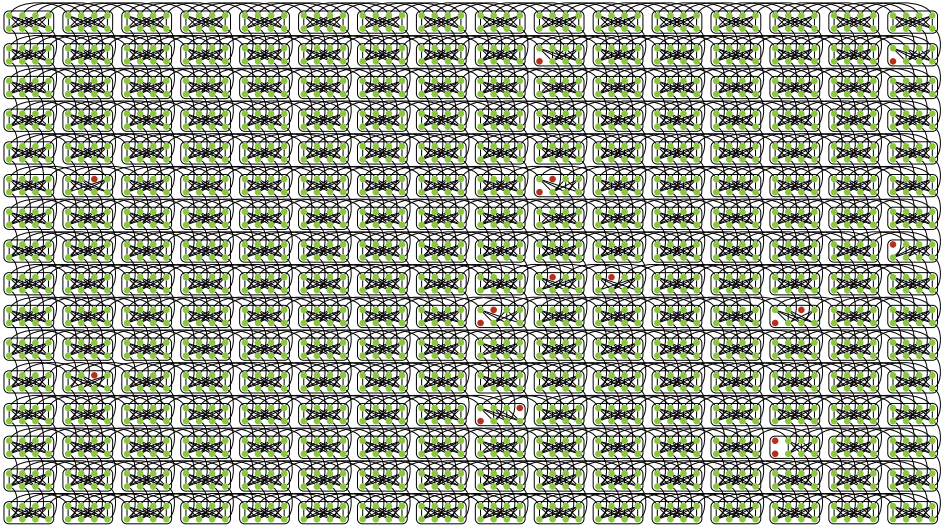} 
   \caption{{The DW2000Q hardware connectivity graph.}  Operational qubits are shown in green, and inoperationable ones are shown in red.  Programmable couplers are shown as black lines connecting the qubits.}
   \label{fig:DW2KQgraph}
\end{figure*}

\begin{figure*}[h!] %
   \centering
   \subfigure[]{\includegraphics[width=0.32\columnwidth]{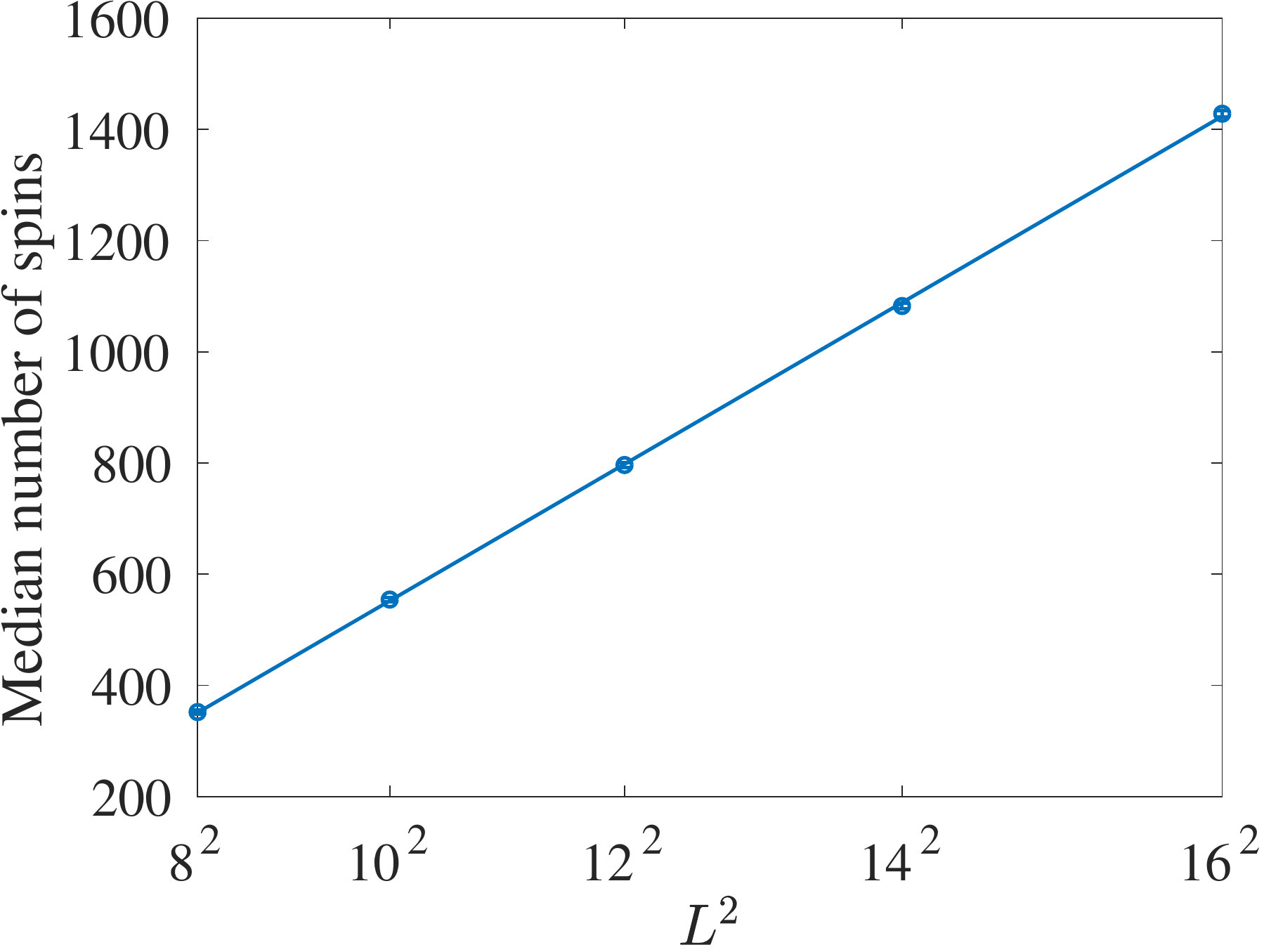} \label{fig:ChimeraSpins}}
   \subfigure[]{\includegraphics[width=0.32\columnwidth]{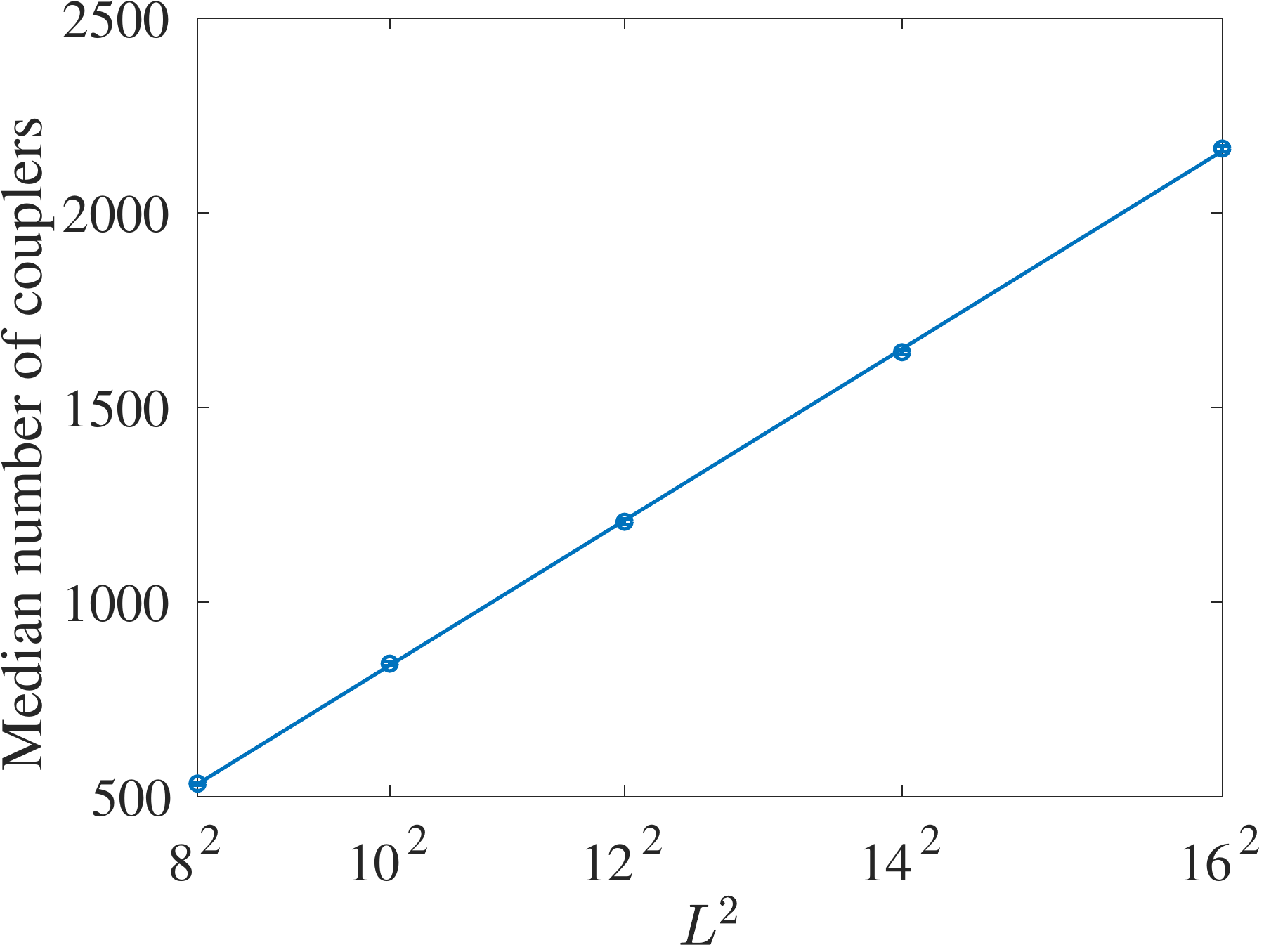} }
   \subfigure[]{\includegraphics[width=0.32\columnwidth]{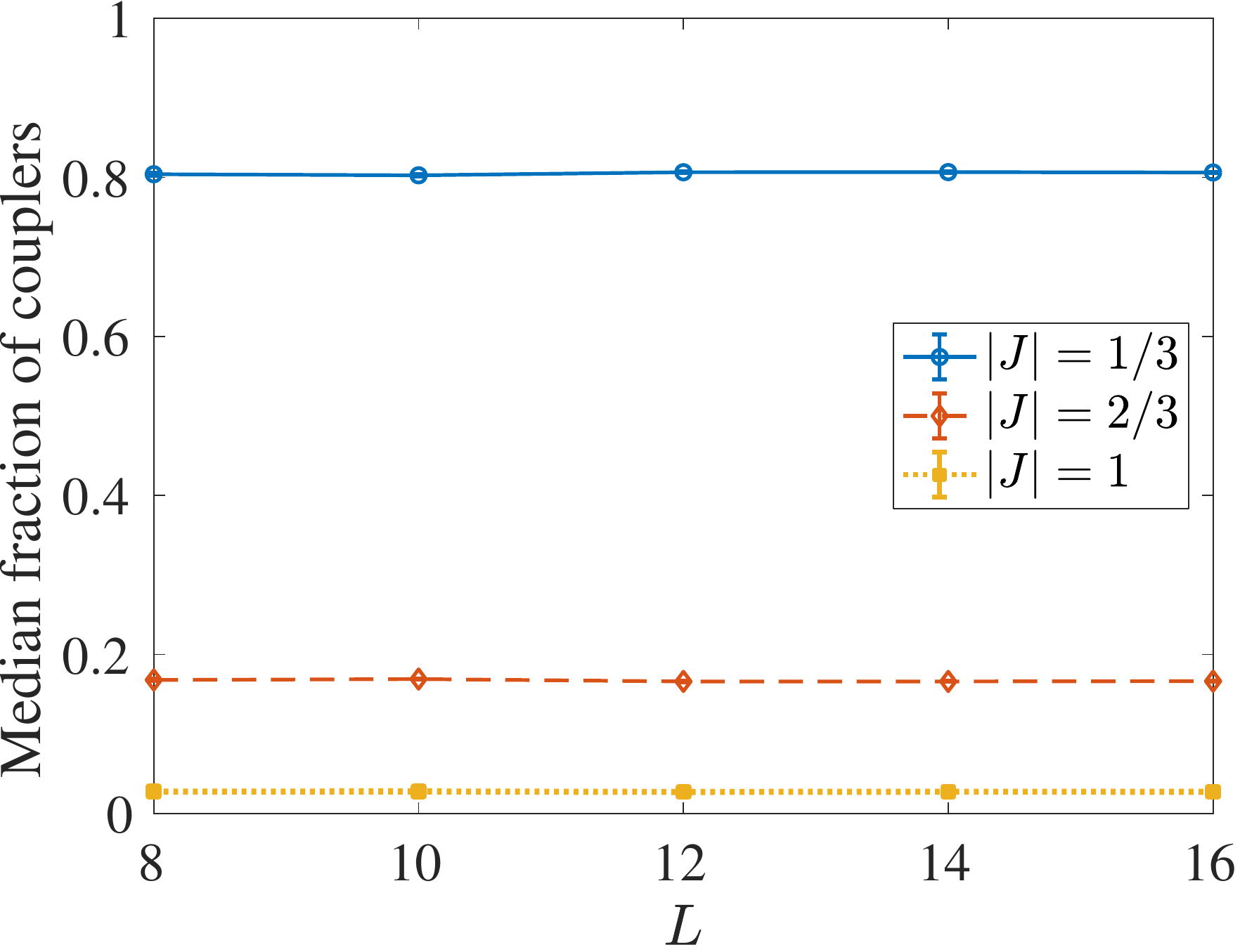} }
   \caption{Properties of the planted-solution instances with two ground states.  (a) Median (over $10^3$ instances) number of spins per instance defined on a $L \times L$ Chimera graph.  Solid line is a linear fit to the data points with a slope of approximately 5.6.  (b) Median (over $10^3$ instances) number of couplers per instance defined on a $L \times L$ Chimera graph.  Solid line is a linear fit to the data points with a slope of approximately 8.5.  (c) Median  (over $10^3$ instances) fraction of the three coupler magnitudes per instance.  Solid lines are meant to guide the eye.  Error bars correspond to two standard deviate error bars  generated using $10^3$ bootstraps over the $10^3$ instances.}
   \label{fig:UniqueGSv6b}
\end{figure*}

\begin{figure*}[h!] %
   \centering
   \subfigure[]{\includegraphics[width=0.32\columnwidth]{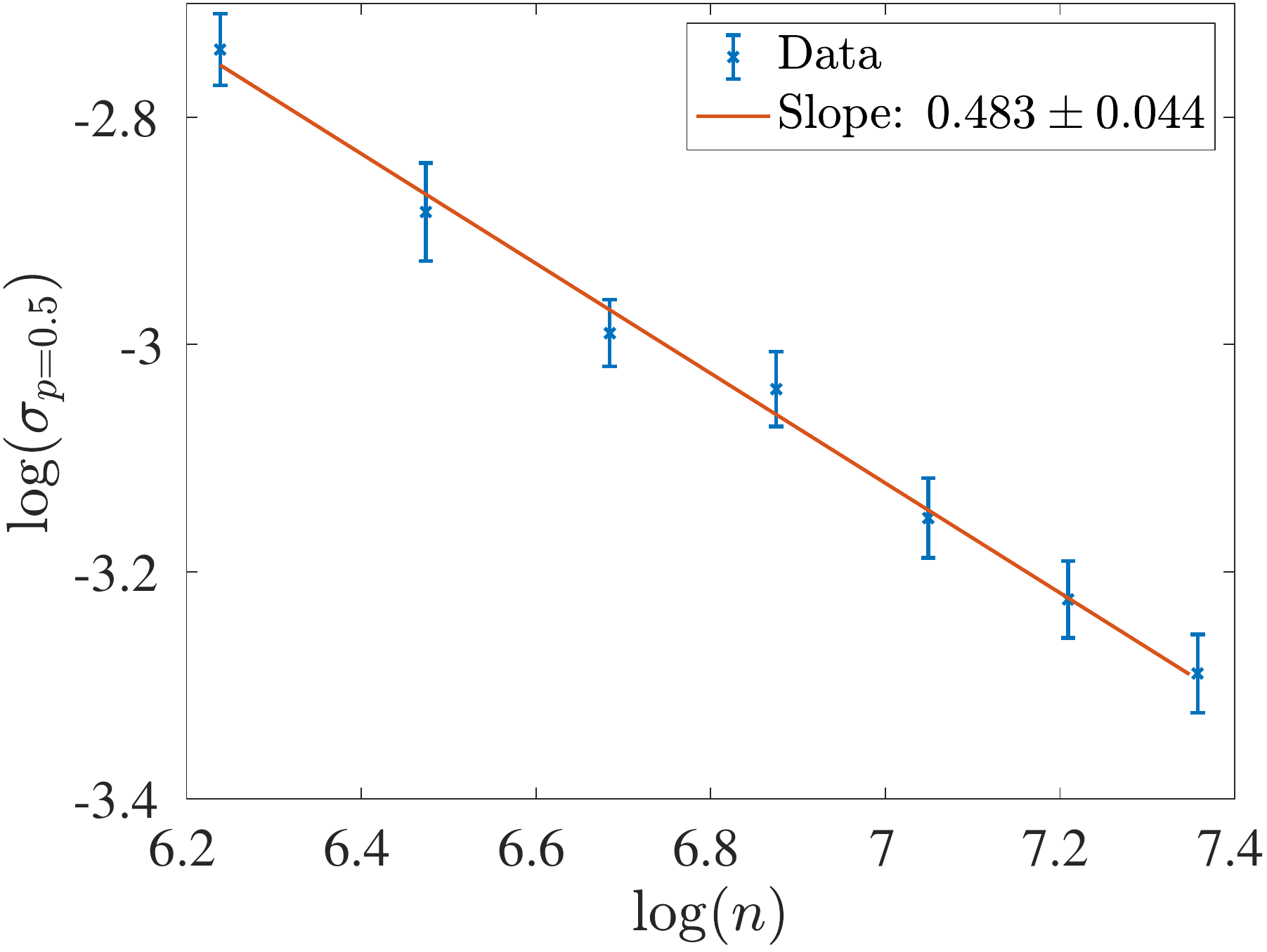} }
   \subfigure[]{\includegraphics[width=0.32\columnwidth]{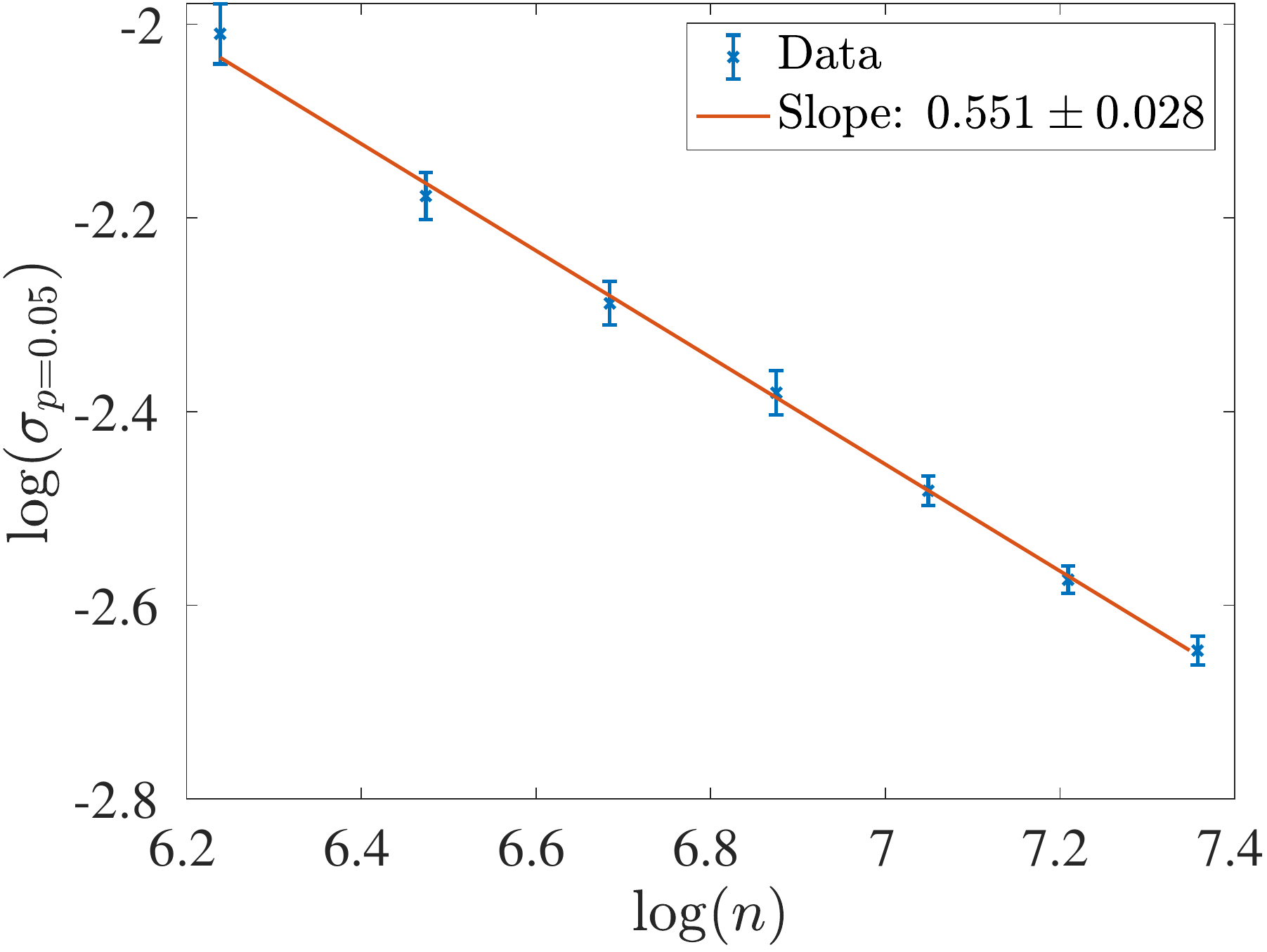} }
   \subfigure[]{\includegraphics[width=0.32\columnwidth]{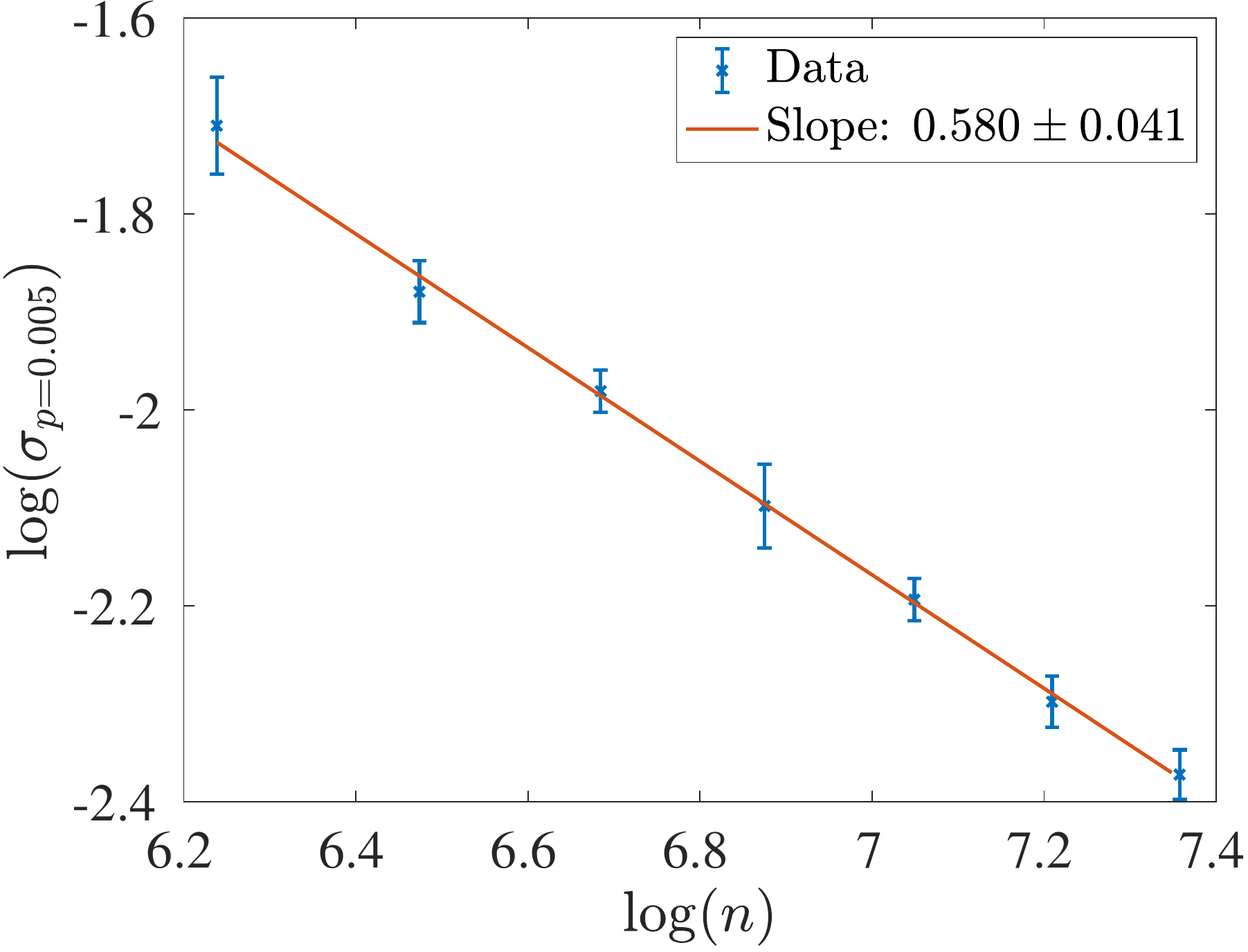} }
   \caption{Examples of our procedure to determine the scaling behavior of $\sigma_p$ with $n$ for the Chimera instances (a) $p=0.5$, (b) $p=0.05$, and (c) $p=0.005$.  In addition to the data points for $\sigma_p$, we show the best fit line through the data and the associated slope.}
   \label{fig:FittingToDataChimera}
\end{figure*}

\begin{figure*}[h] %
   \centering
   \subfigure[]{\includegraphics[width=0.32\columnwidth]{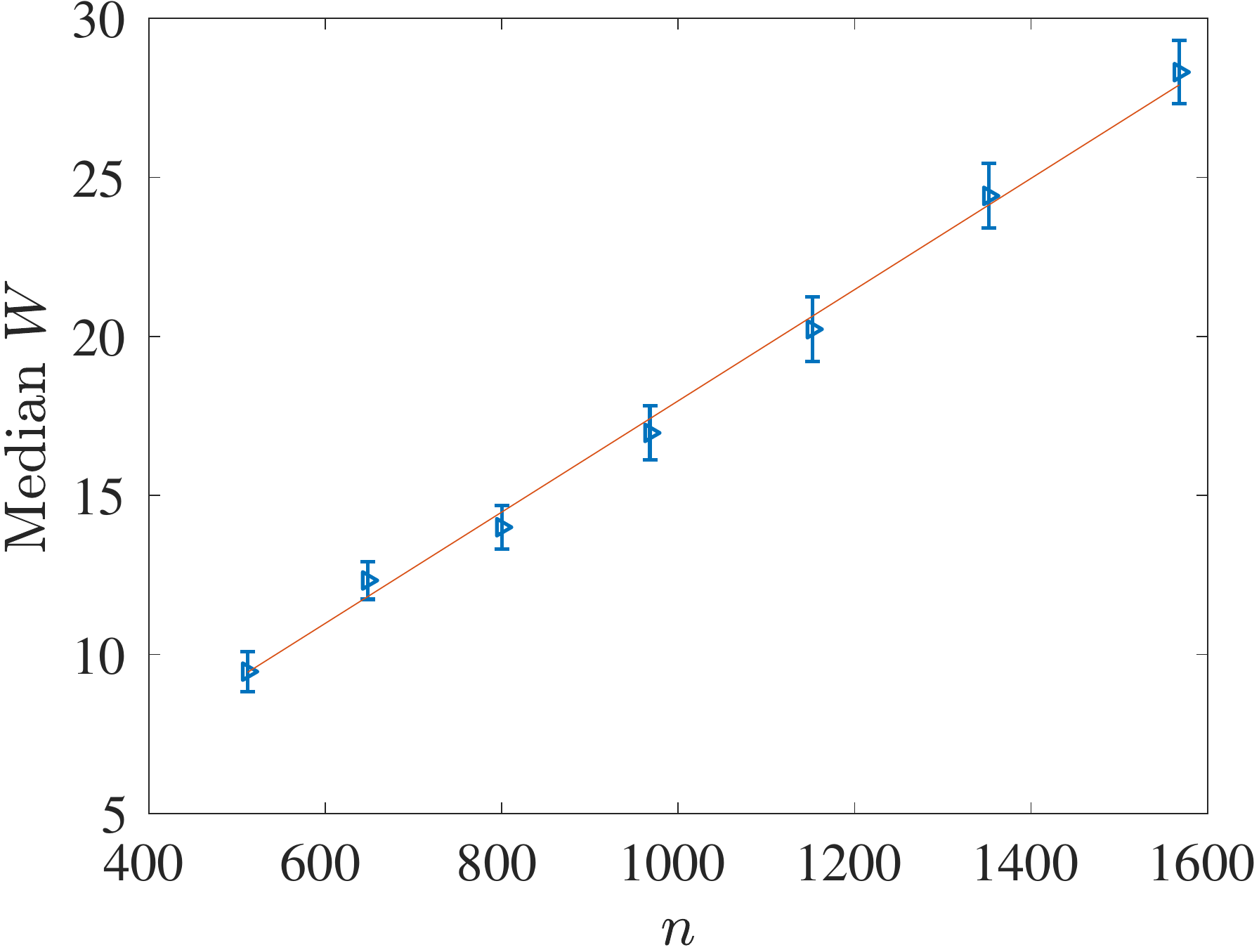} }
      \subfigure[]{\includegraphics[width=0.32\columnwidth]{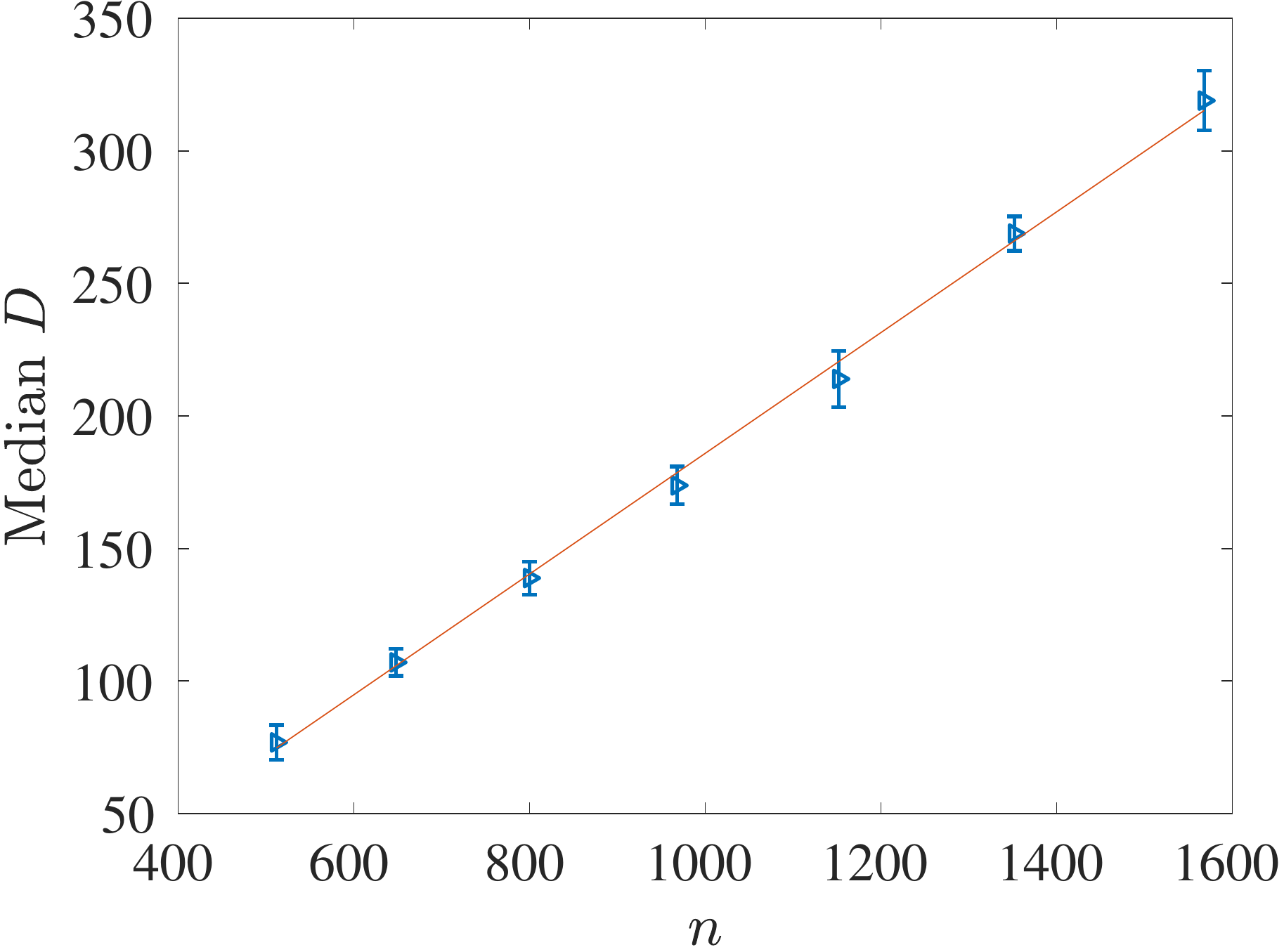} }
      \subfigure[]{\includegraphics[width=0.32\columnwidth]{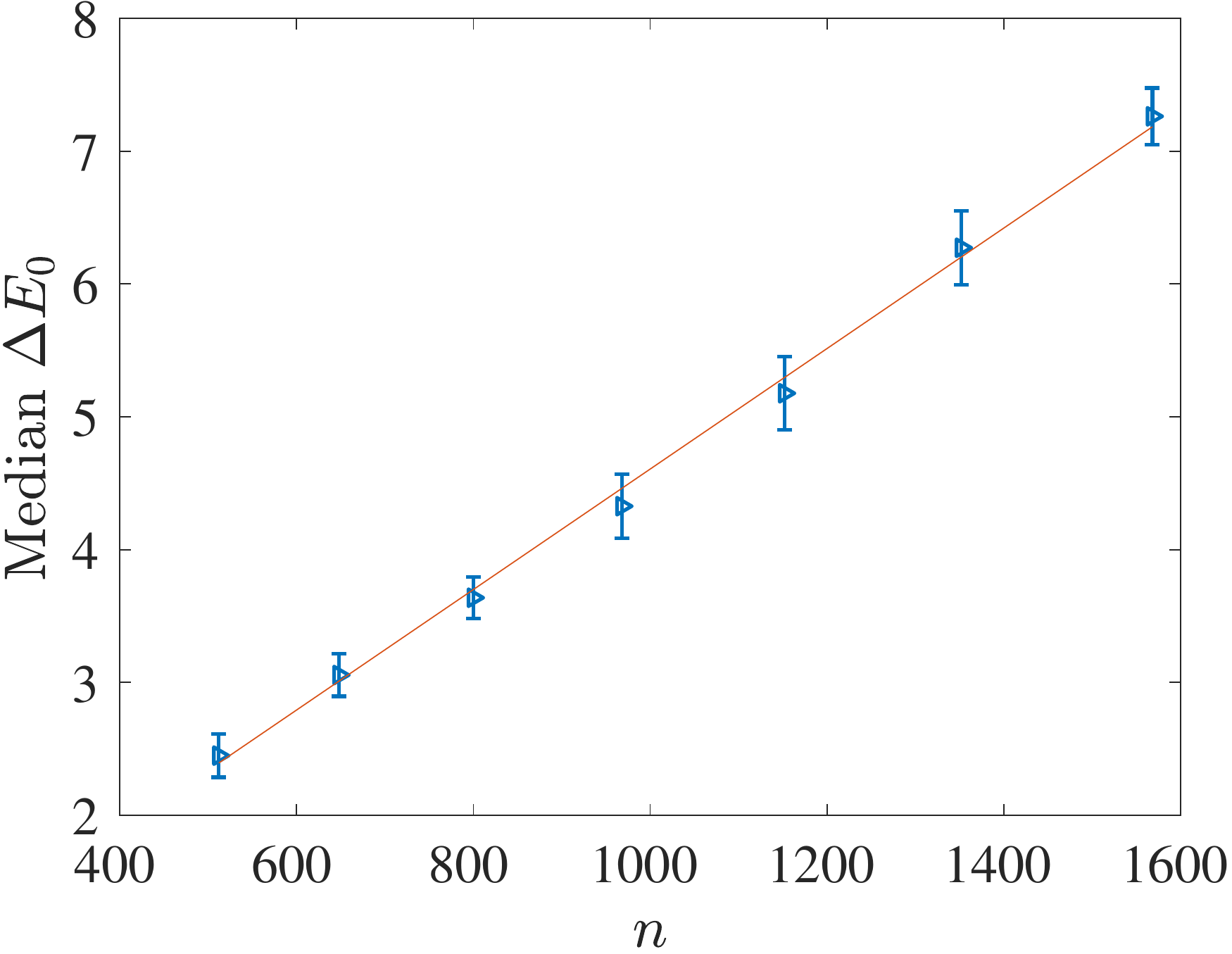} \label{fig:ChimeraMedianEnergy}}
   \caption{Median mass $W$, Hamming distance $D$, and energy from the ground state $\Delta E_0$ (as measured by the intended Hamiltonian) for topologically non-trivial ESs that are more optimal than the GSs for $\sigma = 0.1$. (a) Linear fit given by: $W = 0.487 + 0.017 n$. (b) Linear fit given by  $D =  -41.819 + 0.228 n$. (c) Linear fit given by $\Delta E_0 = 0.071 + 0.005 n$.  Error bars correspond to two standard deviate error bars  obtained from bootstrapping over the 100 instances.}
   \label{fig:Chimera_W_D}
\end{figure*}

\begin{figure*}[h] %
   \centering
   \subfigure[]{\includegraphics[width=0.32\columnwidth]{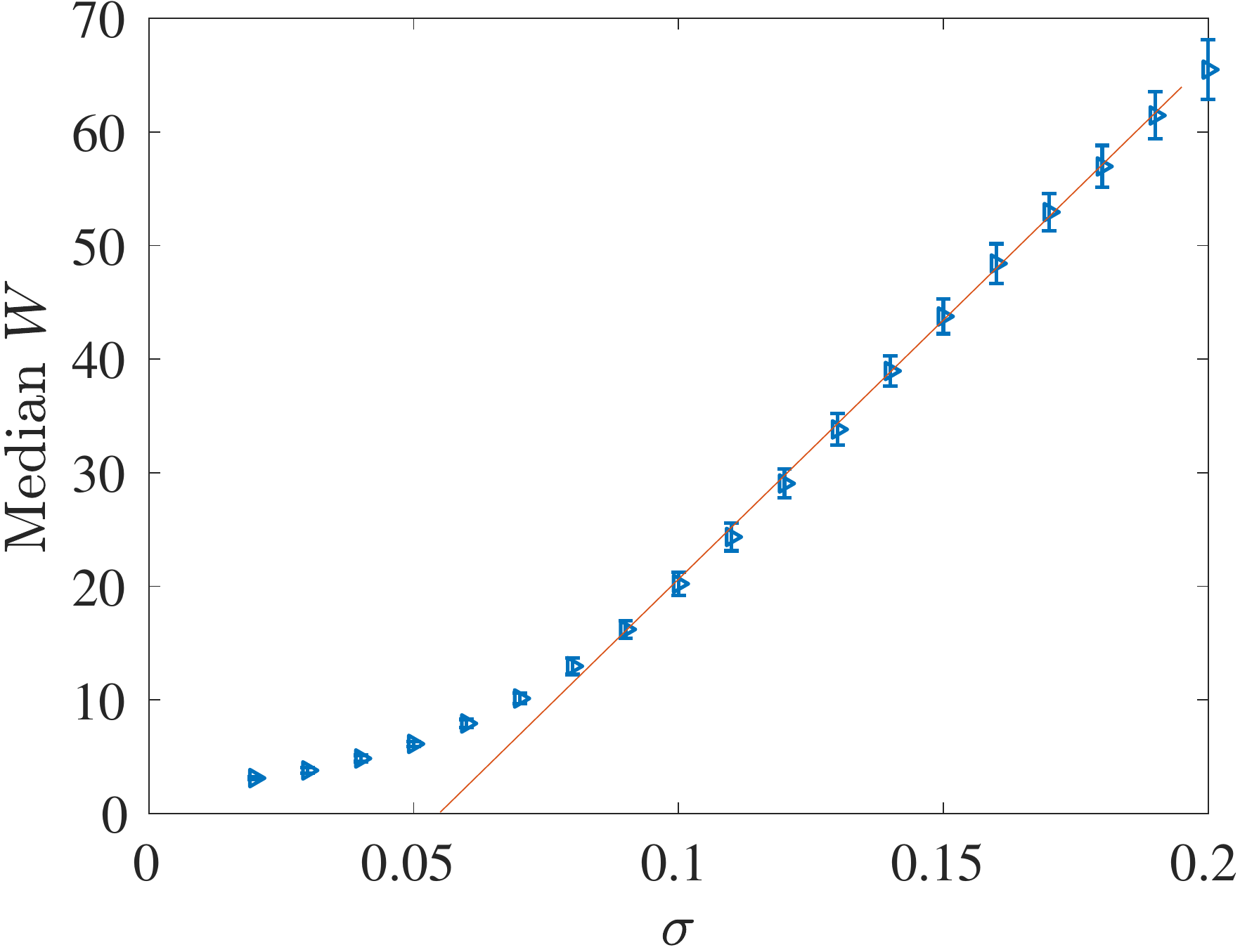} }
      \subfigure[]{\includegraphics[width=0.32\columnwidth]{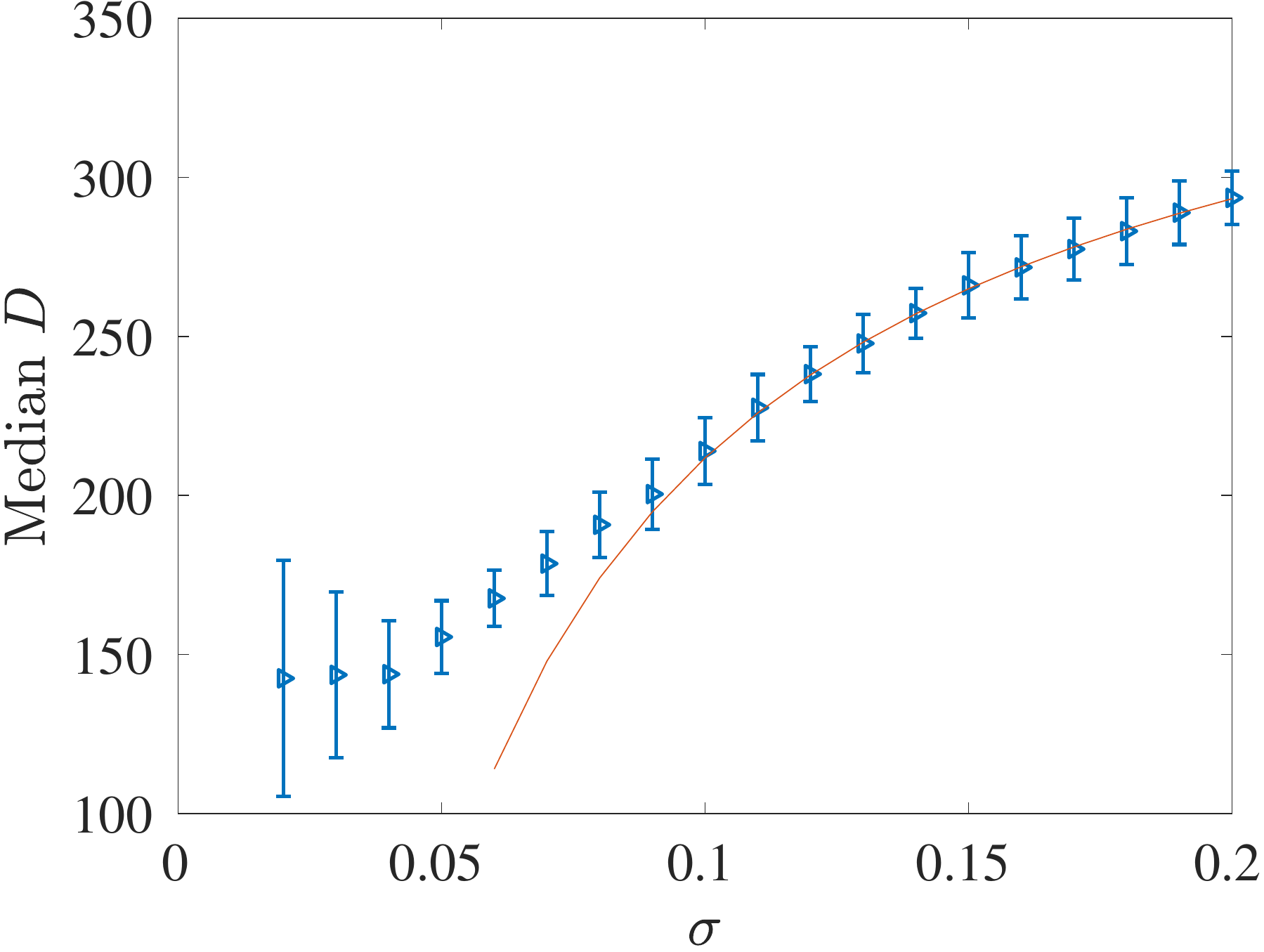} }
      \subfigure[]{\includegraphics[width=0.32\columnwidth]{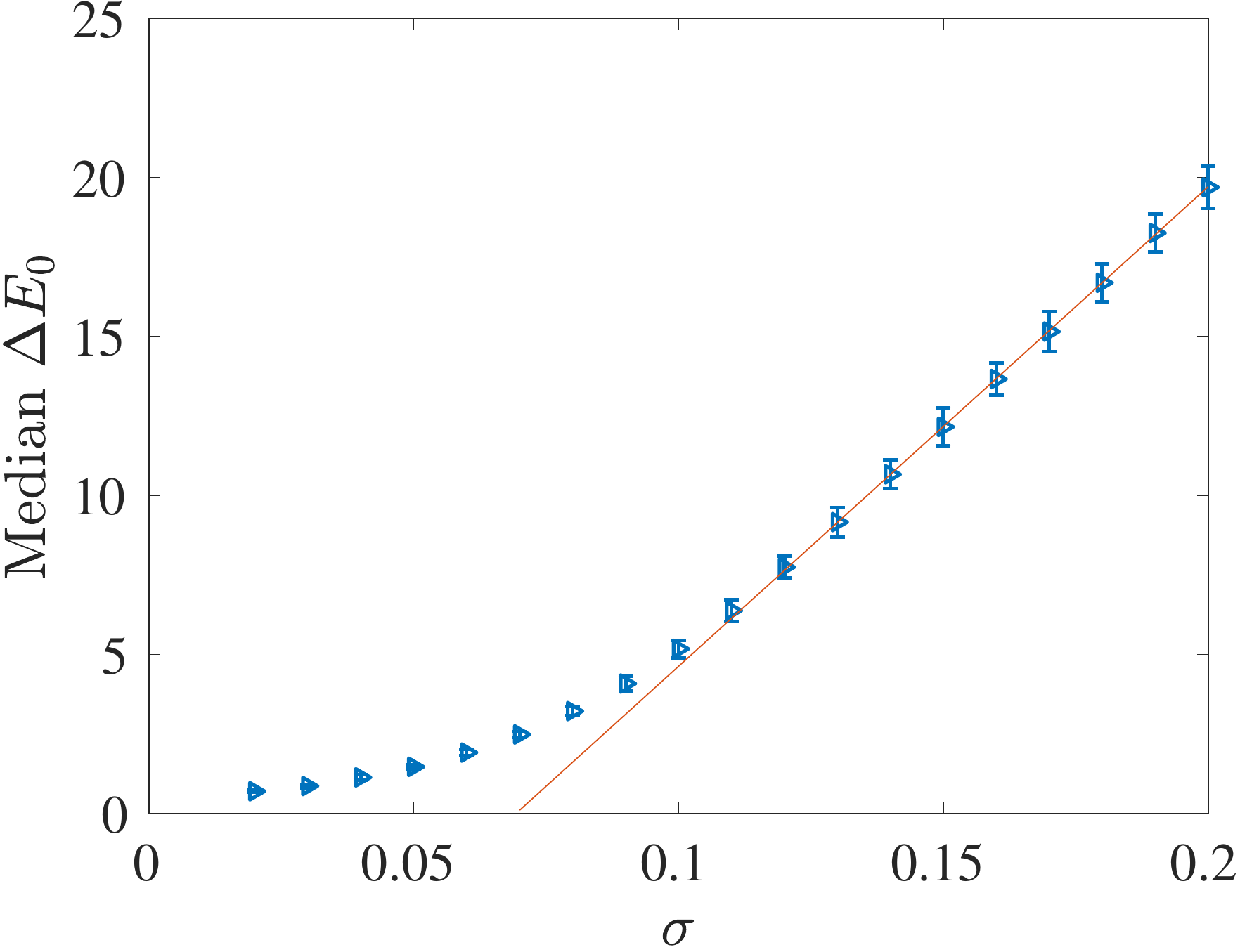} }
   \caption{Median mass $W$, Hamming distance $D$, and energy from the ground state $\Delta E_0$ (as measured by the intended Hamiltonian) for topologically non-trivial ESs that are more optimal than the GSs for instances defined on a $12\times 12$ Chimera grid. (a) Linear fit given by: $W = -31.34 + 452.29 \sigma$. (b) Fit given by  $D =  399.61 - 28.38\sigma^{-0.82}$. (c) Linear fit given by $\Delta E_0 = -10.45 +  150.76 \sigma$.  Error bars correspond to two standard deviate error bars  obtained from bootstrapping over the 100 instances.}
   \label{fig:Chimera_W_D2}
\end{figure*}

\begin{figure*}[h] %
   \centering
   {\includegraphics[width=0.32\columnwidth]{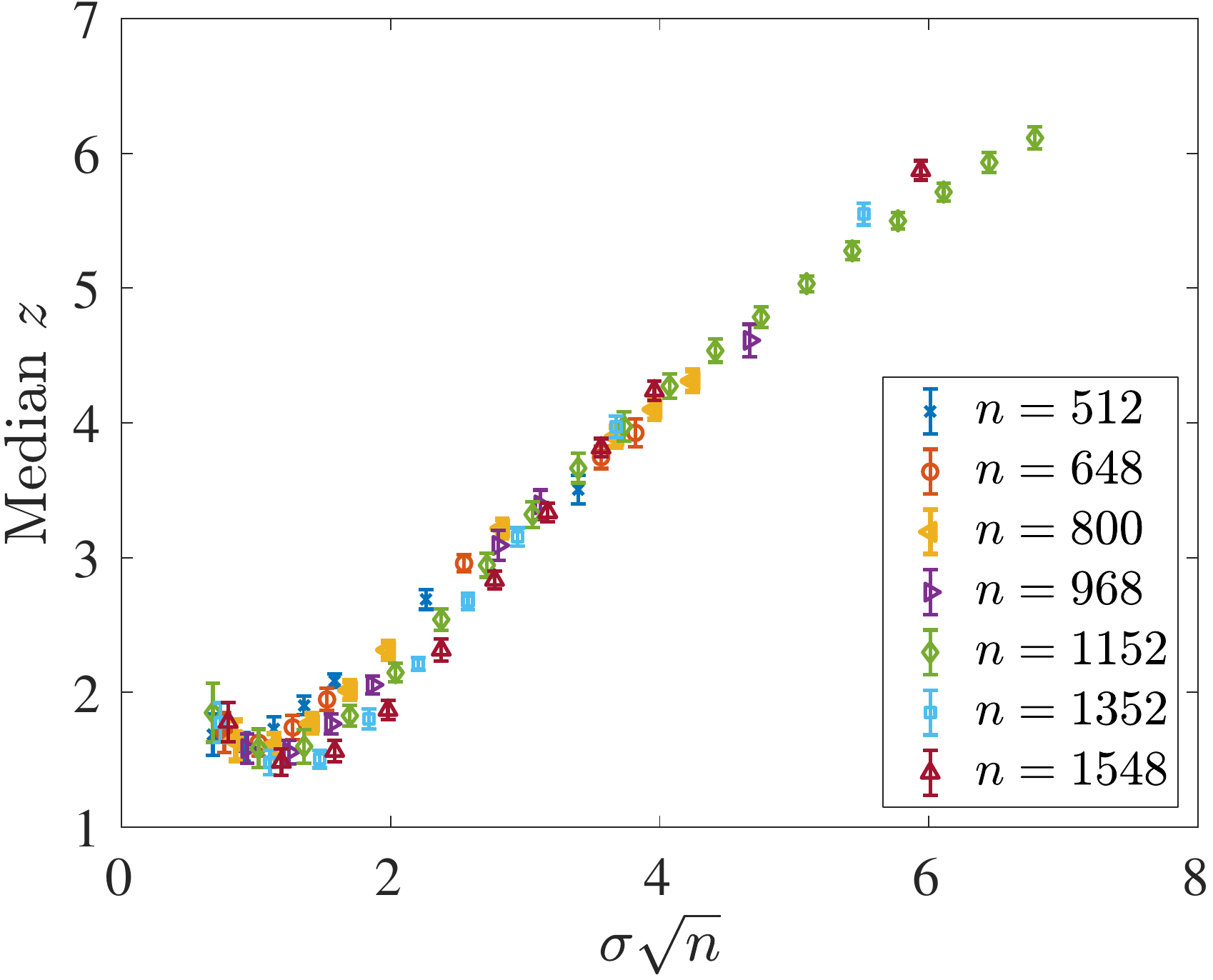} }
   \caption{The median $z$ value for topologically non-trivial ESs that are more optimal than the GSs for instances defined on a $L\times L$ Chimera grid with $n = 8L^2$.}
   \label{fig:Chimera_z}
\end{figure*}

\begin{figure*}[t]
\begin{center}
\includegraphics[width=.32\columnwidth]{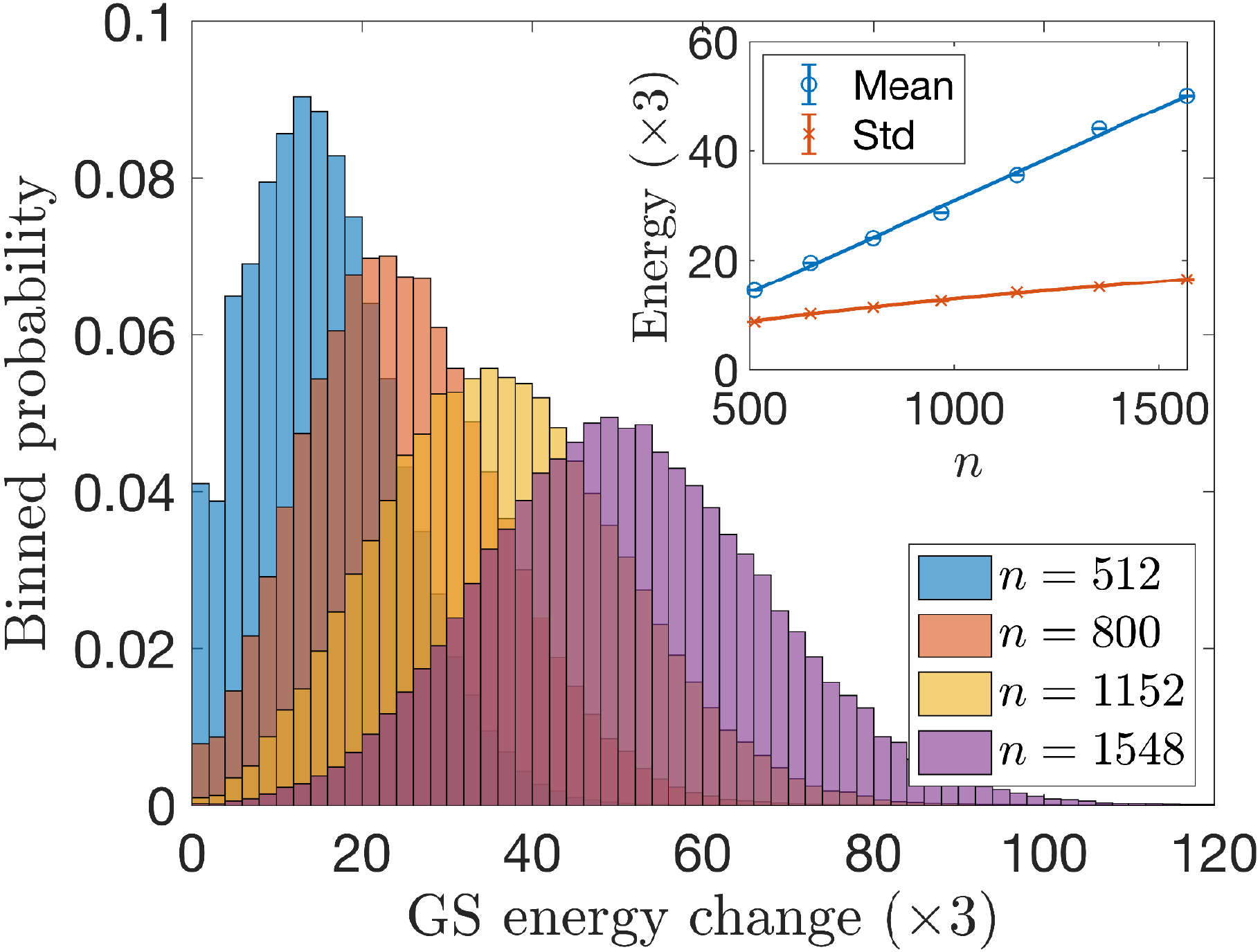}
\caption{{Histogram of the residual energy of the GSs of the implemented (noisy) Hamiltonian as measured by the intended Hamiltonian.}  Here we use the planted-solution Chimera instances defined on varying grid sizes $L \in [8,14]$ with a noise level of $\sigma = 0.15$.  Only the topologically non-trivial ESs of the intended Hamiltonian are shown.  For each problem size, we use $10^3$ noise realization for each of the $10^2$ instances, so each problem size histogram includes $10^5$ data points. Inset: the scaling of the mean and standard deviation of the distribution of GS energies.  The solid curves are fits to $\mu_E = a + n b$ and $\sigma_E = a + \sqrt{n} b$ with $a = -2.88, b = 0.034$ and $a = -1.37, b = 0.45$ respectively with $n= 8L^2$.
}
\label{fig:enDist2}
\end{center}
\end{figure*}

The results shown in Fig.~3 of the main text were taken on a 4th generation D-Wave processor, the D-Wave 2000Q  (DW2000Q) `Whistler' processor, installed in the NASA Advanced Supercomputing Facility at NASA's Ames Research Center.  The processor connectivity is given by a $16 \times 16$ grid of unit cells, where each unit cell is composed of 8 qubits with a $K_{4,4}$ bipartite connectivity, forming the `Chimera' graph~\cite{Choi1,Choi2}.  Of the total 2048 qubits on the device, 2031 are operational qubits.  The hardware connectivity graph of the processor is illustrated in Fig.~\ref{fig:DW2KQgraph}.  

The planted-solution instance studied on the DW2000Q processor is defined identically as described above except the frustrated loops are restricted to a $12\times12$ subgraph of the $16\times16$ graph of the device.  Because of the allowed programming range of the device, no coupling can have magnitude greater than 1, so before submission to the processor, the couplings of the instance are normalized so that the largest coupling strength is 0.5.  We then introduce gaussian noise with mean 0 and standard deviation $\sigma$ to both the couplings and local fields.  If any coupling or local field has magnitude larger than 1, we set them to 1.  The annealing time used was $40 \mu s$.

\section{3-regular 3-XORSAT instances} \label{app:XORSAT}

Here we discuss another class of examples we have studied, which is not of the Ising type, but the theory presented in the main text can be easily generalized to include this case.  The intended Hamiltonian can be written in the generic form:
\beq
H = \sum_{\langle i,j,k \rangle}  \sigma_i^z \sigma_j^z \sigma_k^z
\eeq
We consider the case of 3-regular 3-XORSAT, where for each spin we randomly pick three other spins to which to couple.  All couplings are picked to be antiferromagnetic with strength $1$.  Because all terms in the Hamiltonian are of the form $+\sigma_i^z \sigma_j^z \sigma_k^z$, the ground state is simply that all-spins-down state.  The implemented Hamiltonian we consider has the form:
\beq
\tilde{H} = \sum_{i} \delta h_i \sigma_i^z  + \sum_{\langle i,j,k \rangle} \left( 1 + \delta J_{ijk} \right) \sigma_i^z \sigma_j^z \sigma_k^z
\eeq
where as in the main text we take $\delta h_i$ and $\delta J_{ijk}$ to be $\mathcal{N}(0,\sigma^2)$.  

In order to find the ground state of the implemented Hamiltonian, we used parallel tempering \cite{Geyer:91,Hukushima:1996} with iso-energetic Houdayer cluster updates \cite{Houdayer,PhysRevLett.115.077201}.  Our implementation used 32 inverse-temperatures distributed geometrically between a maximum inverse-temperature of 20 and a minimum inverse-temperature of 0.1.  We ran the simulations for $2\times 10^6$  Monte Carlo sweeps, with Houdayer cluster updates and temperature swaps occurring after every 10 sweeps.  
 The lowest energy configuration found during the simulation is reported.  If this state has an energy (according to the implemented Hamiltonian) lower than the all-down spin state, then we are guaranteed that the identity of the ground state has changed for the implemented Hamiltonian.

\begin{figure*}[h!] %
   \centering
   \subfigure[]{\includegraphics[width=0.32\columnwidth]{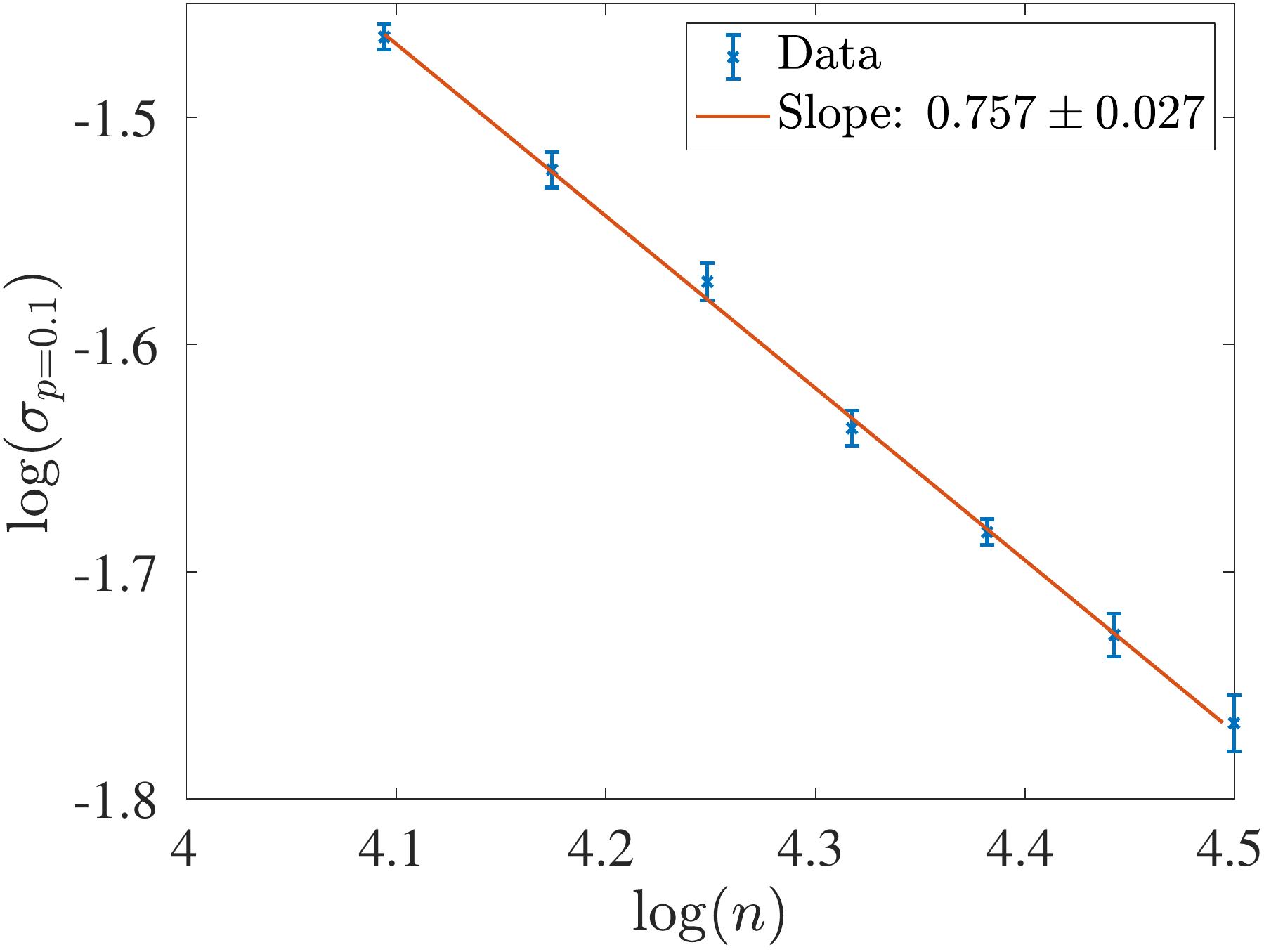} }
   \subfigure[]{\includegraphics[width=0.32\columnwidth]{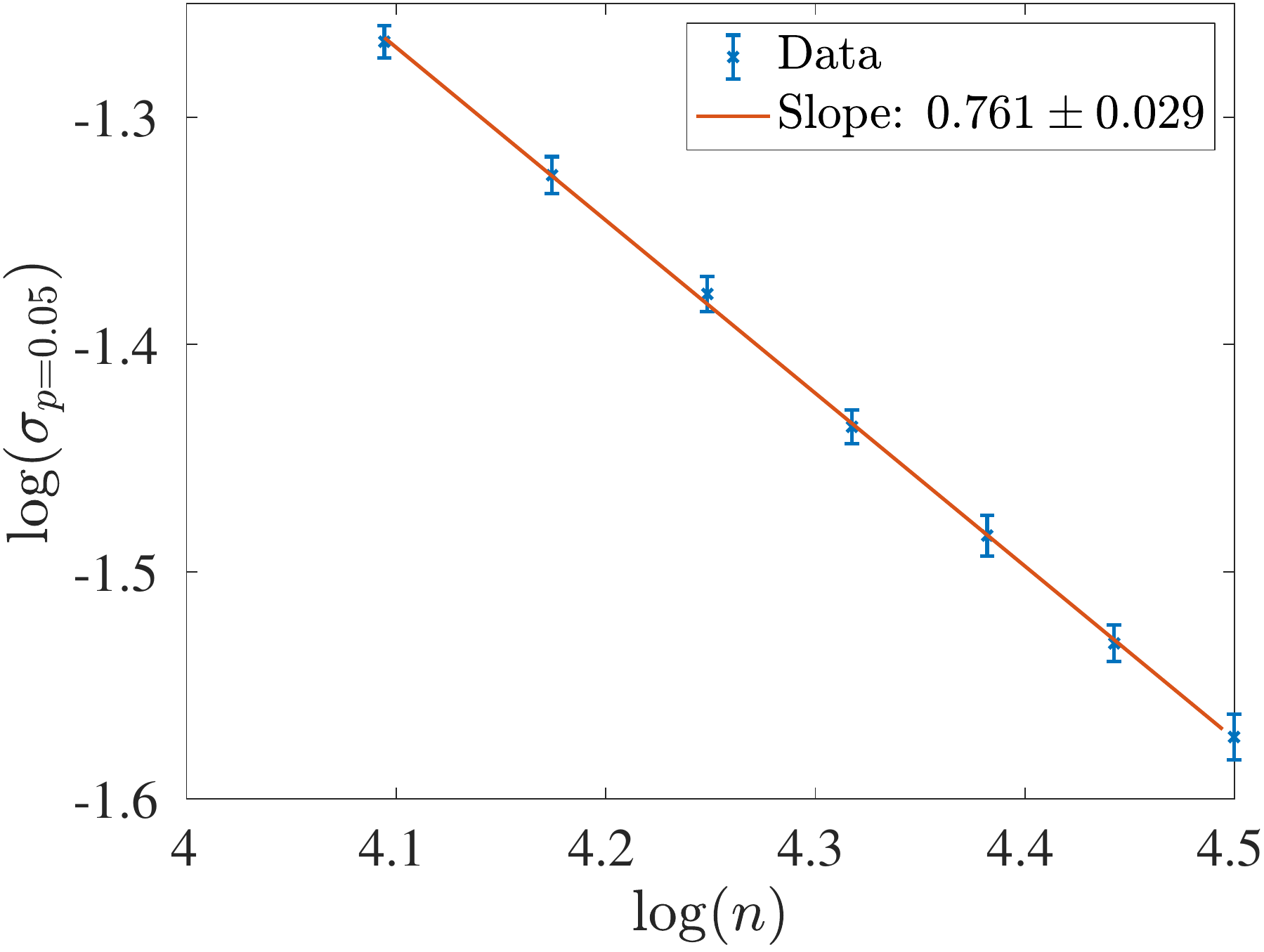} }
   \subfigure[]{\includegraphics[width=0.32\columnwidth]{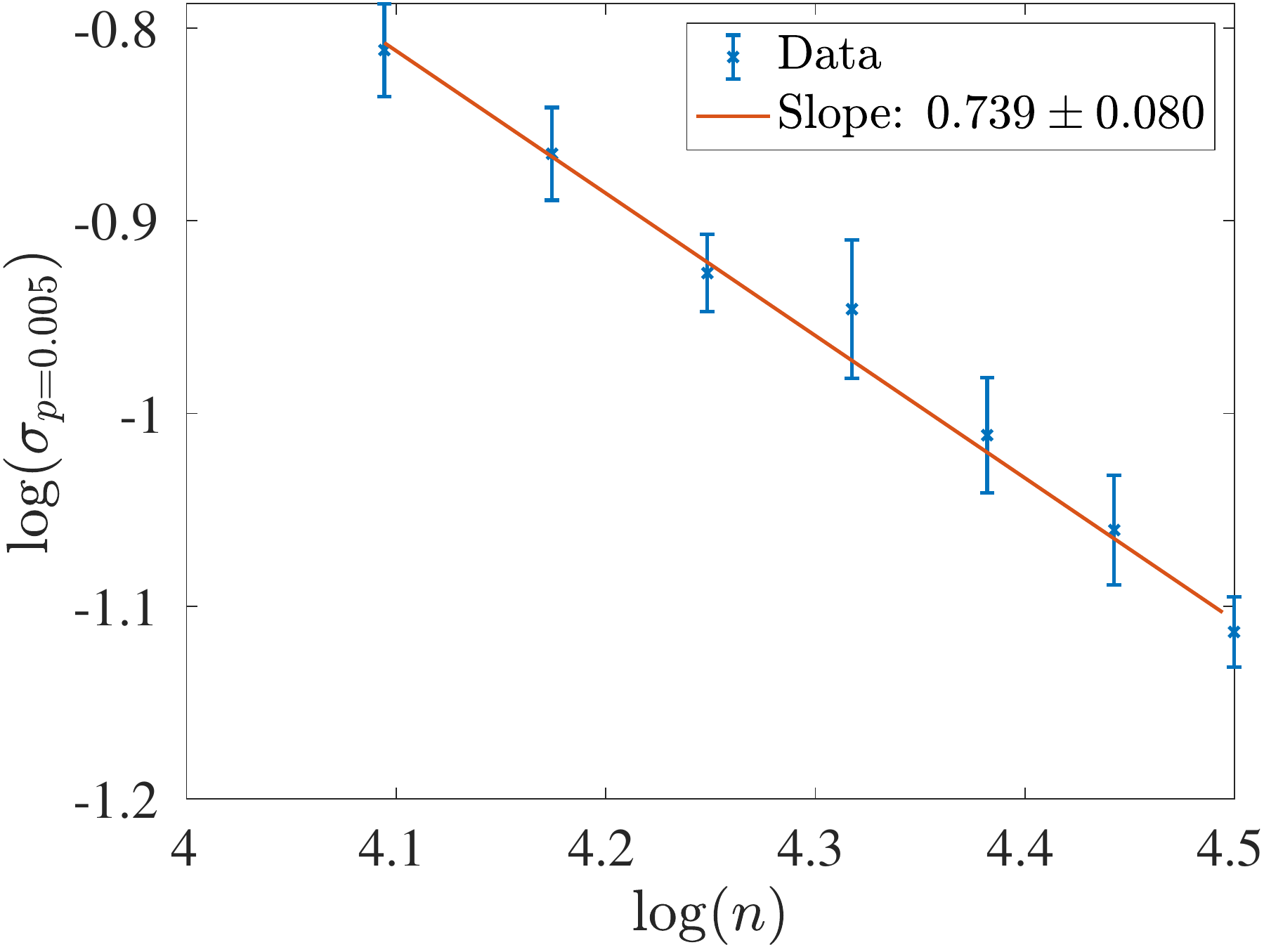} }
   \caption{Examples of our procedure to determine the scaling behavior of $\sigma_p$ with $n$ for the 3-XORSAT instances for (a) $p=0.1$, (b) $p=0.05$, and (c) $p=0.005$.  In addition to the data points for $\sigma_p$, we show the best fit line through the data and the associated slope.}
   \label{fig:FittingToDataXOR}
\end{figure*}

\begin{figure*}[t] %
   \centering
\subfigure[]{\includegraphics[width=0.32\columnwidth]{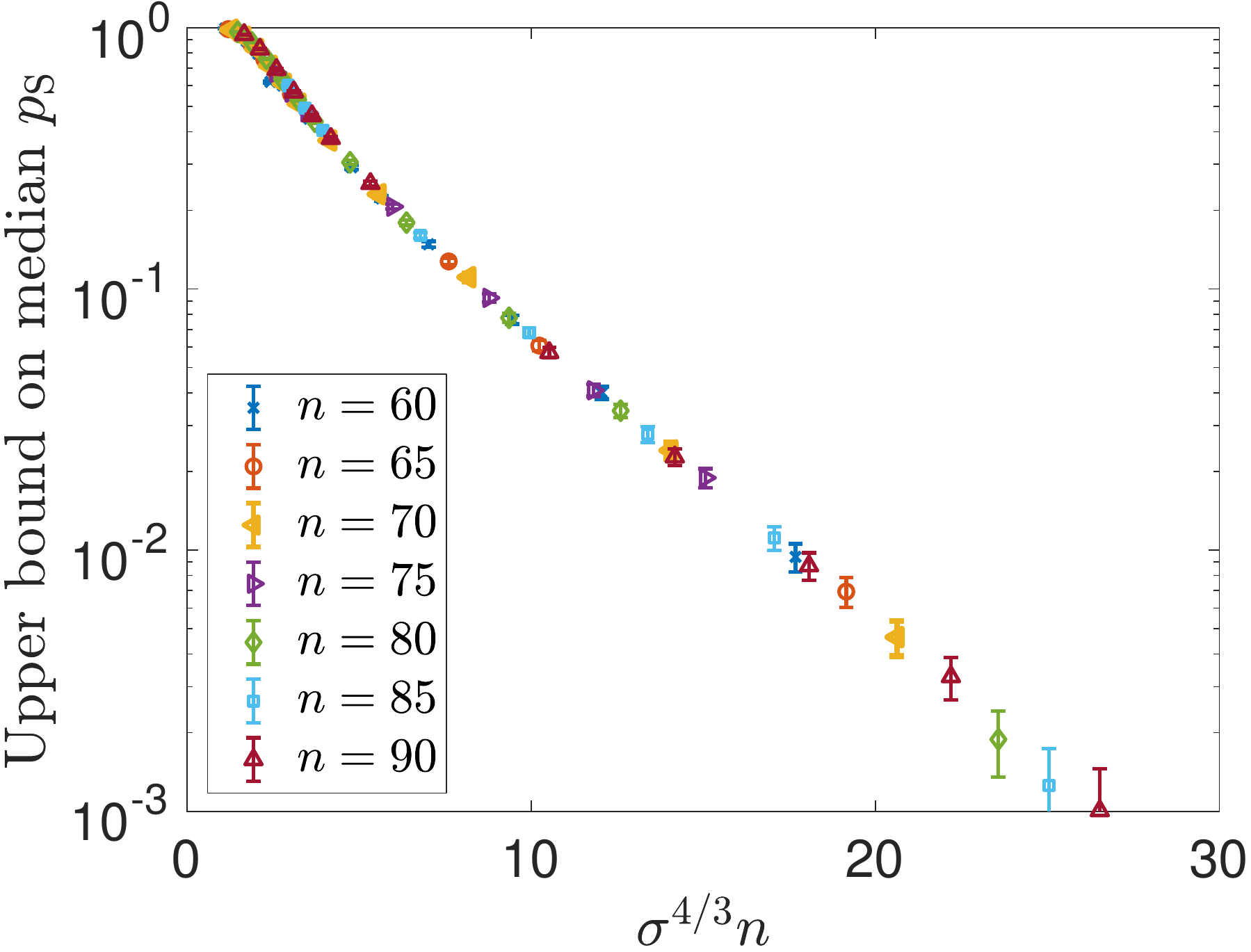} }
\subfigure[]{\includegraphics[width=0.32\columnwidth]{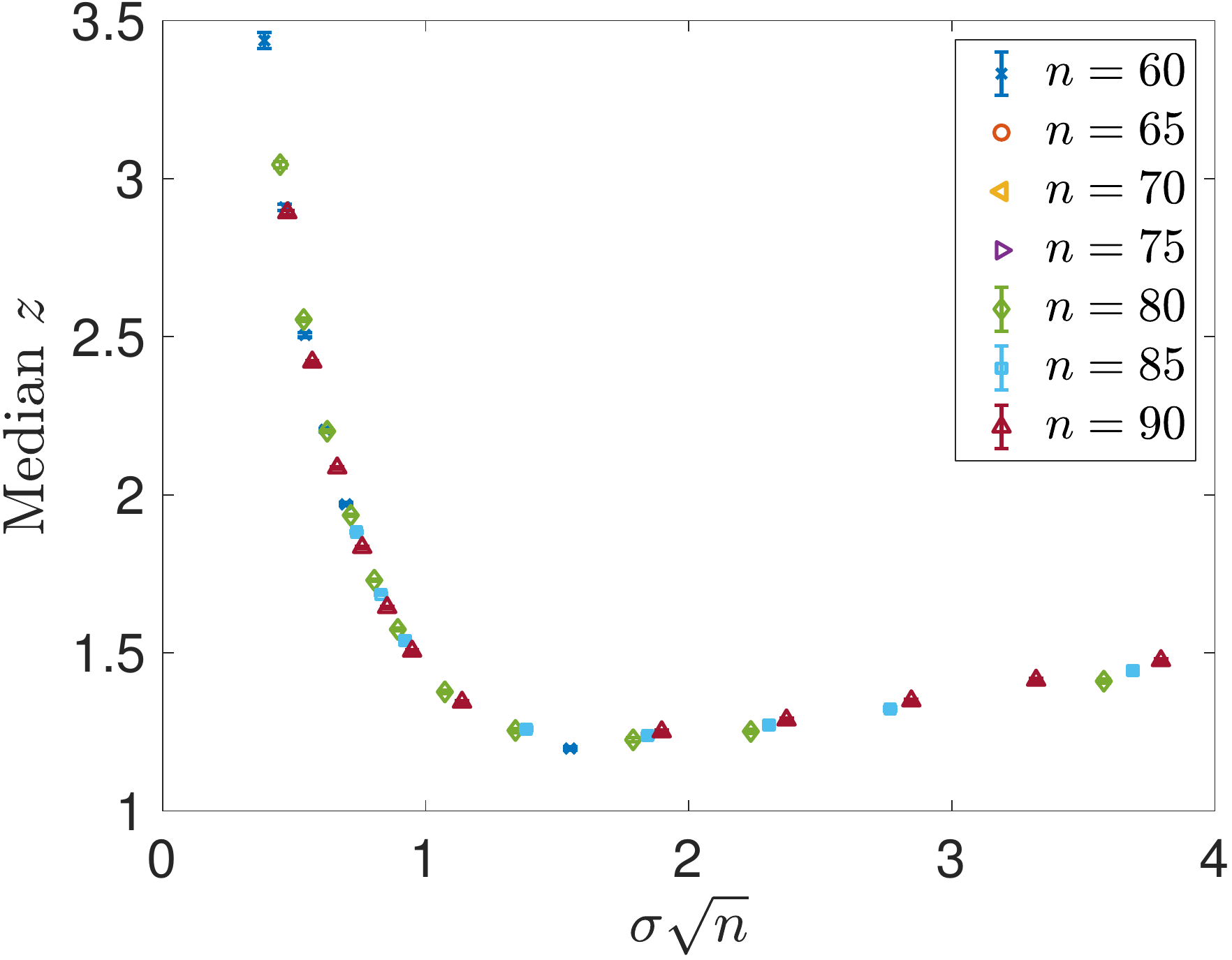}}
      \caption{(a) Upperbound on the median probability $p_{\mathrm{S}}$ that the ground state of the intended Hamiltonian remains the ground state of the implemented Hamiltonian for varying planted-solution instances on 3-regular 3-XORSAT sizes $n \in \left[ 60, 90 \right]$ and Gaussian noise strengths $\sigma \in \left[ 0.02,0.4\right]$.  For each instance, $p_{\mathrm{S}}$ is determined by averaging over 1000 noise realizations, and the median is taken over 100 instances. (b) The median $z$ value for topologically non-trivial ESs that are more optimal than the GSs.  Error bars correspond to two standard deviate error bars  obtained from bootstrapping over the 100 instances.}
   \label{fig:XOR_Collapse}
\end{figure*}

\begin{figure*}[t] %
   \centering
   \subfigure[]{\includegraphics[width=0.32\columnwidth]{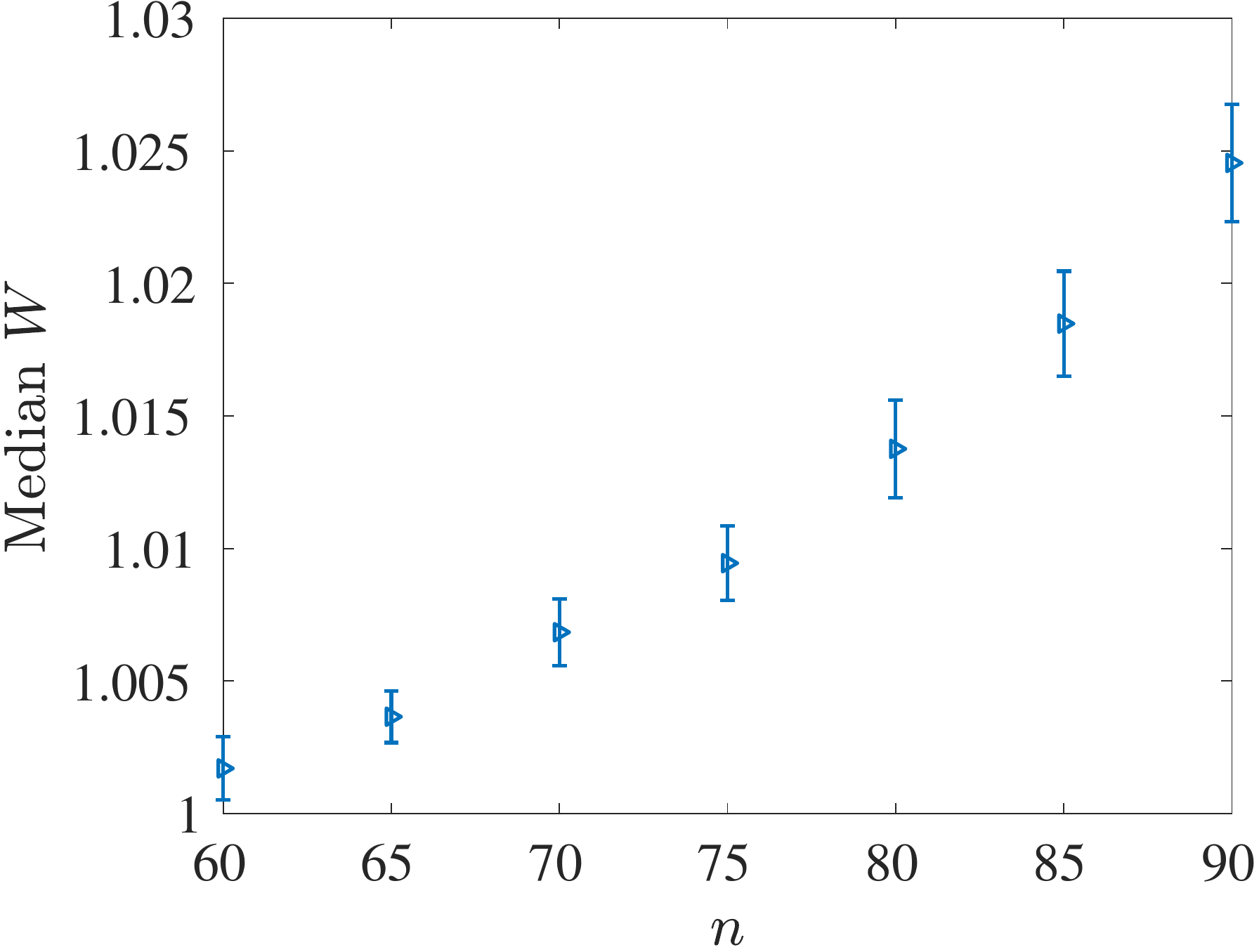}}
      \subfigure[]{\includegraphics[width=0.32\columnwidth]{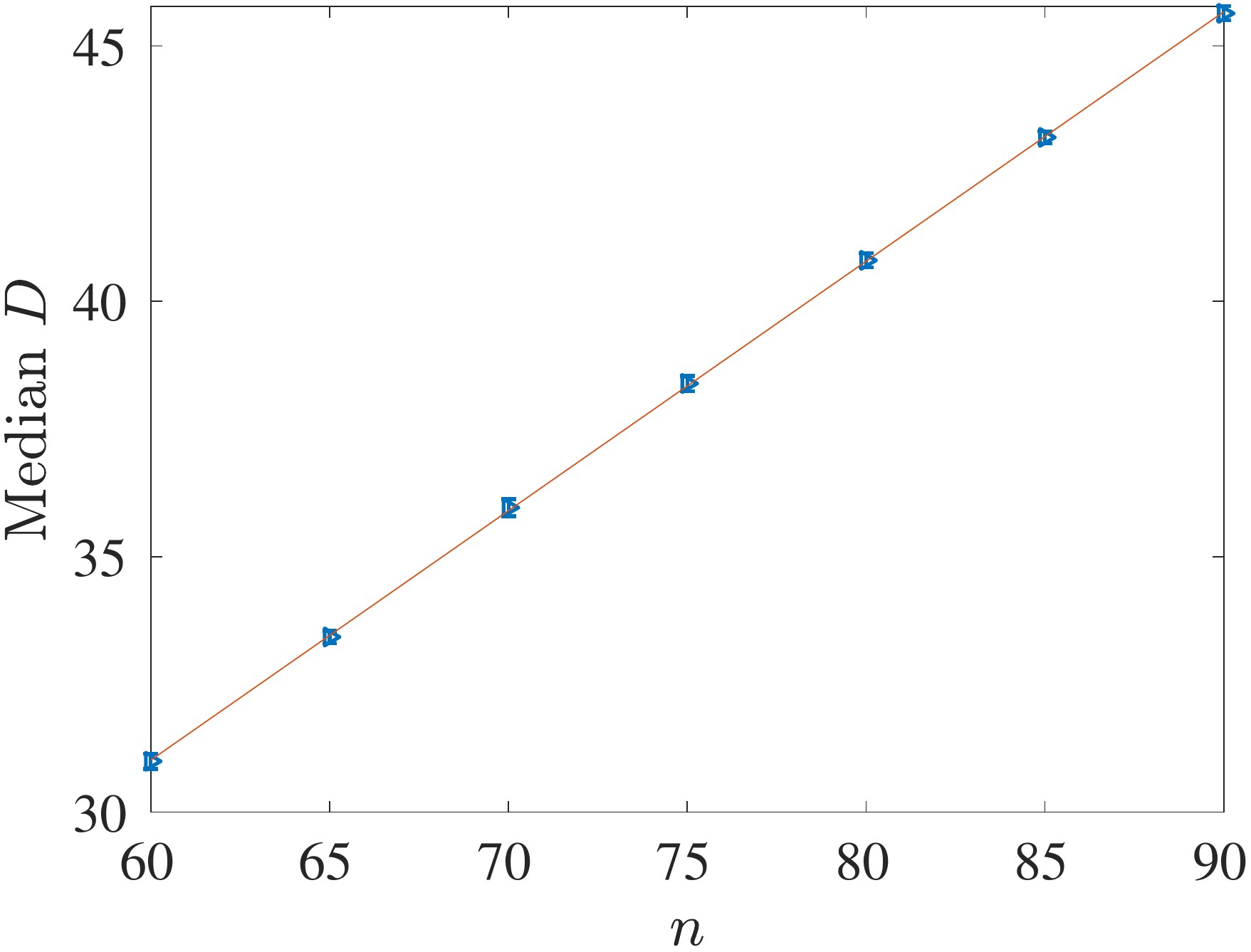} \label{fig:XORMedianHD}}
            \subfigure[]{\includegraphics[width=0.32\columnwidth]{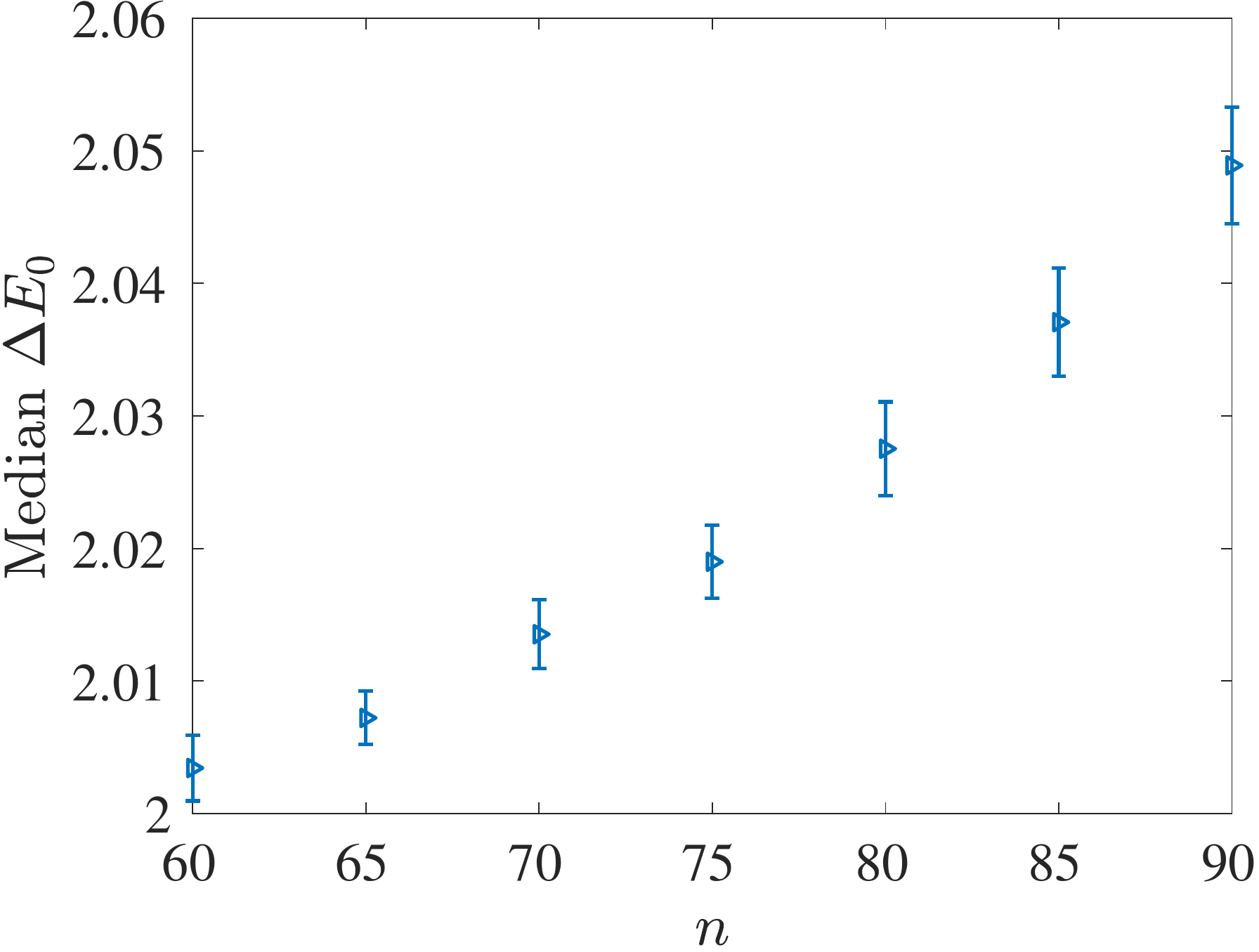} \label{fig:XORMedianEnergy}}
      \caption{(a) Median mass $W$, (b) median Hamming distance $D$, and (c) median energy from GS (as measured by the intended Hamiltonian) for topologically non-trivial ESs that are more optimal than the GSs for $\sigma = 0.1$. For (b) the linear fit given by: $W = 1.73 + 0.49 n$.}
   \label{fig:XOR_W_D}
\end{figure*}

We show our results in Figs.~\ref{fig:FittingToDataXOR}, \ref{fig:XOR_Collapse}, and \ref{fig:XOR_W_D}.  The collapse of the data suggests that the noise level must be scaled as $1/n^{2/3}$, which is different from the Chimera example presented in the main text, and is closer to the worse case estimate given in the main text's Eq.~(10).  The behavior of the median $z$ is again such that $z \gtrsim 1$ over the entire range of parameters we studied.  

\section{Square grid instances} \label{app:SquareGrid}
Here we discuss a class of Ising instances defined on a square grid with random $\pm 1$ instances.  In order to find a ground state of the intended Hamiltonian, we map the Ising Hamiltonian to the minimum-weight perfect-matching (MWPM) problem~\cite{Bieche:1980} and run the Blossom V algorithm~\cite{Kolmogorov2009} to determine the solution to the MWPM problem.  In order to use the same algorithm for the (noisy) implemented Hamiltonians, we restrict our analysis only to perturbations of the couplings. 

\begin{figure*}[h!] %
   \centering
   \subfigure[]{\includegraphics[width=0.32\columnwidth]{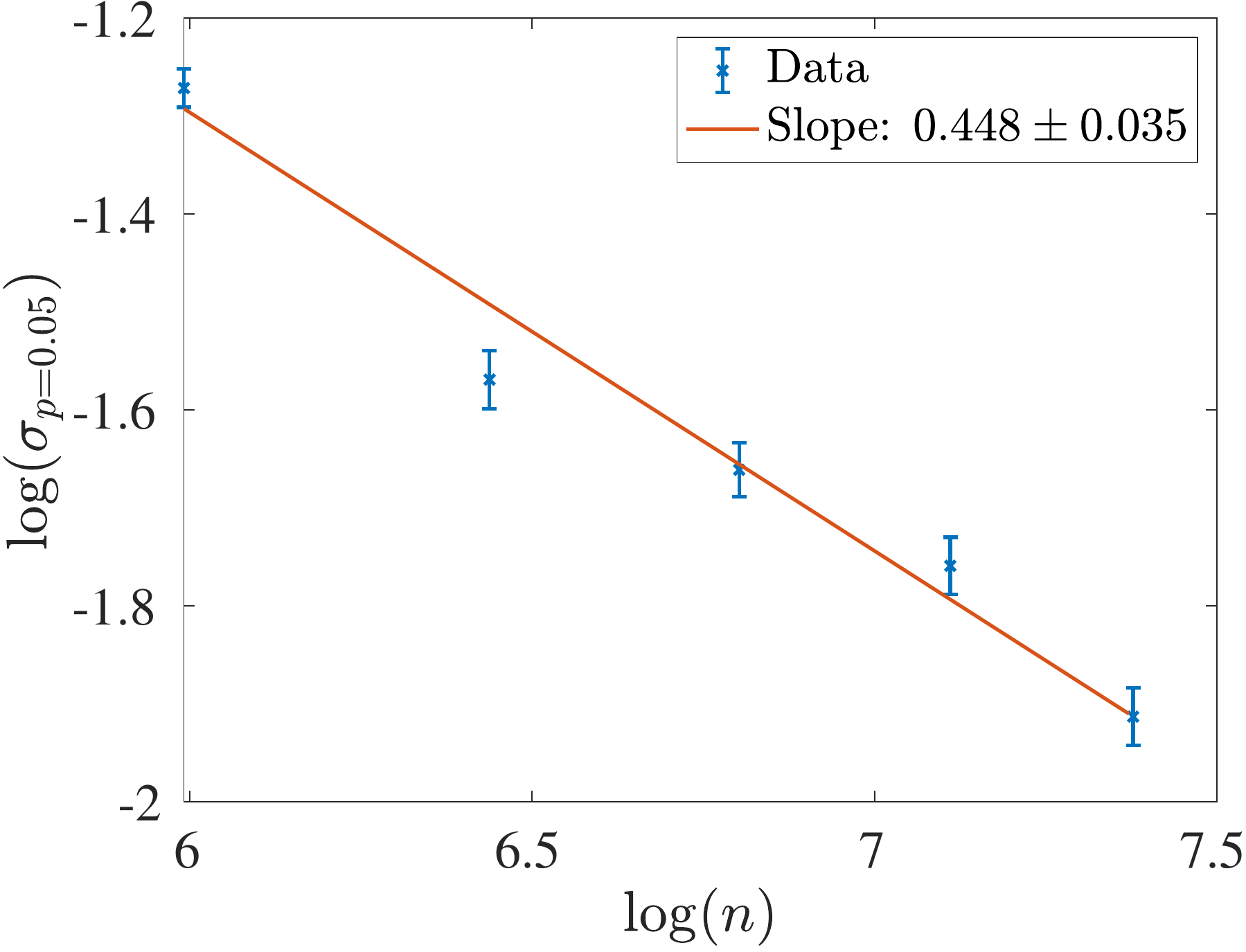} }
   \subfigure[]{\includegraphics[width=0.32\columnwidth]{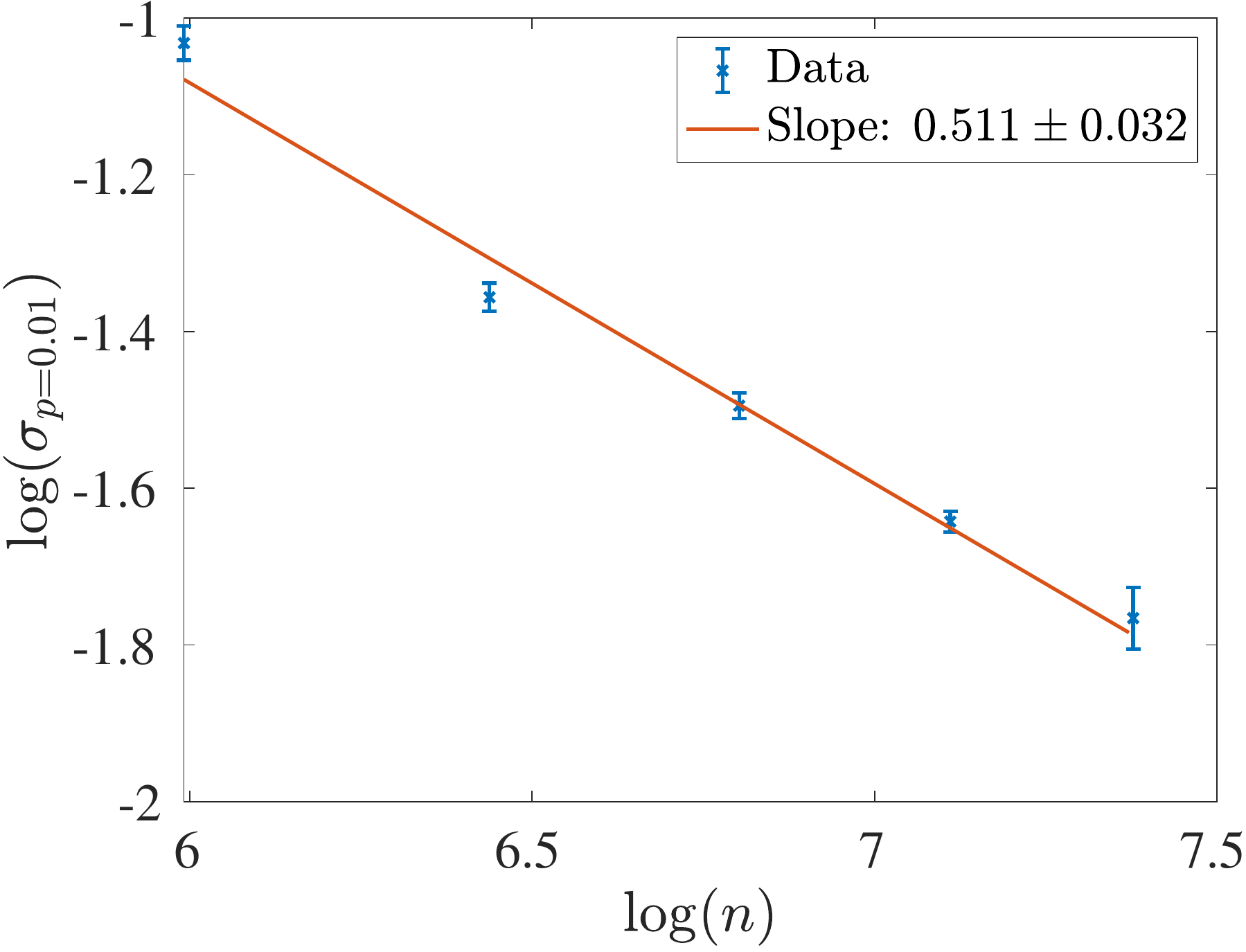} }
   \subfigure[]{\includegraphics[width=0.32\columnwidth]{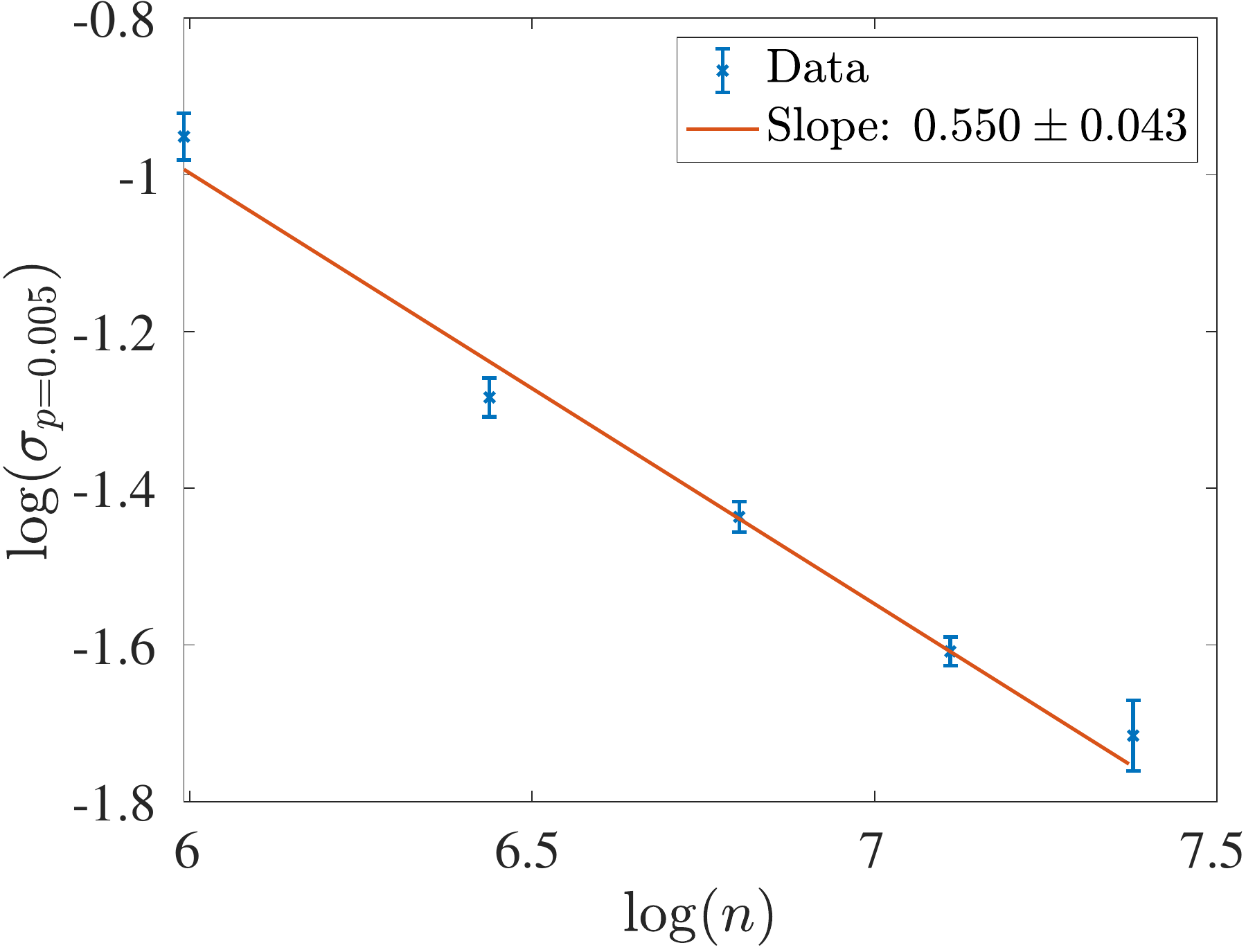} }
   \caption{Examples of our procedure to determine the scaling behavior of $\sigma_p$ with $n$ for the square grid instances for (a) $p=0.05$, (b) $p=0.01$, and (c) $p=0.005$.  In addition to the data points for $\sigma_p$, we show the best fit line through the data and the associated slope.}
   \label{fig:FittingToDataSquare}
\end{figure*}

\begin{figure*}[t] %
   \centering
      \subfigure[]{\includegraphics[width=0.32\columnwidth]{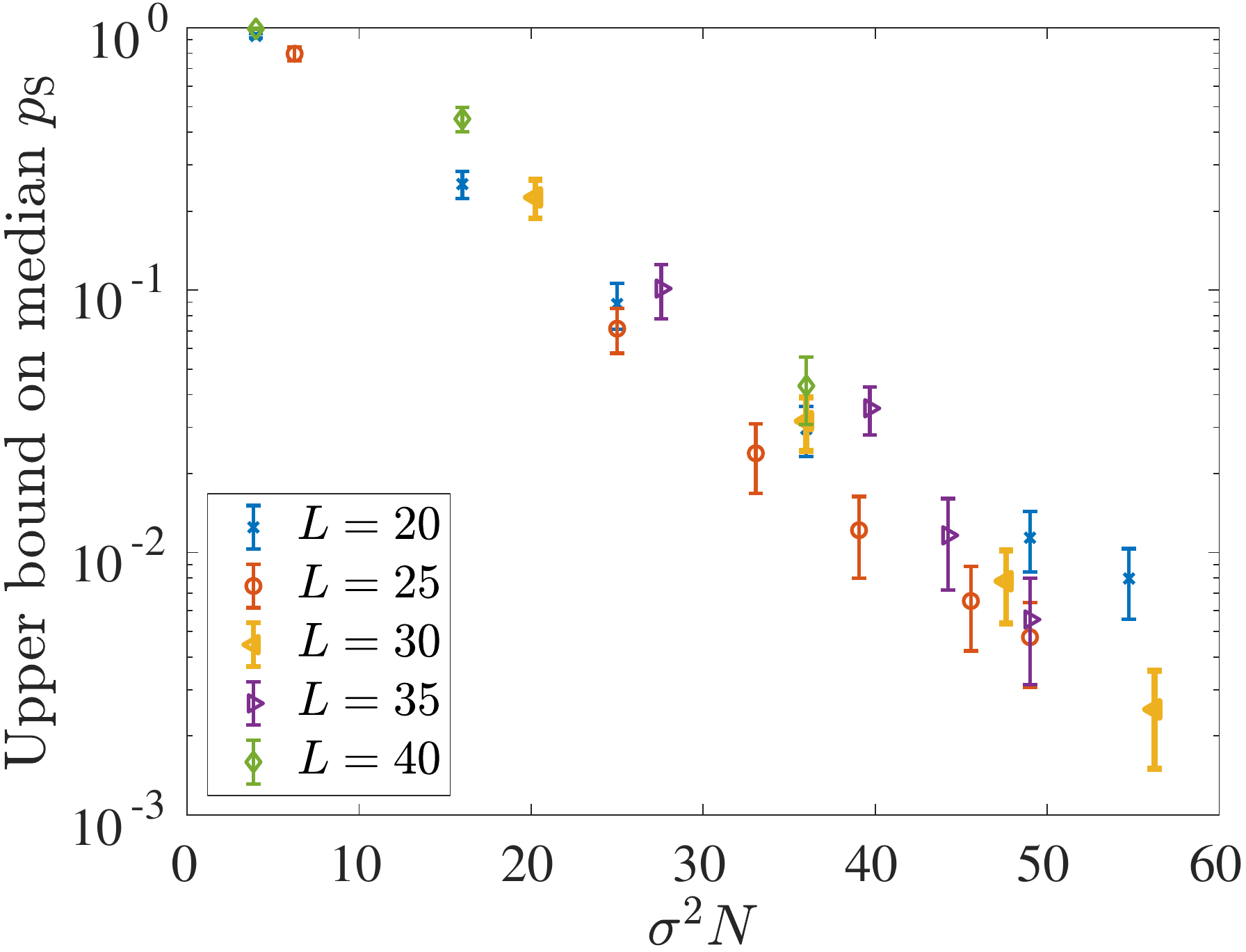} }
            \subfigure[]{\includegraphics[width=0.32\columnwidth]{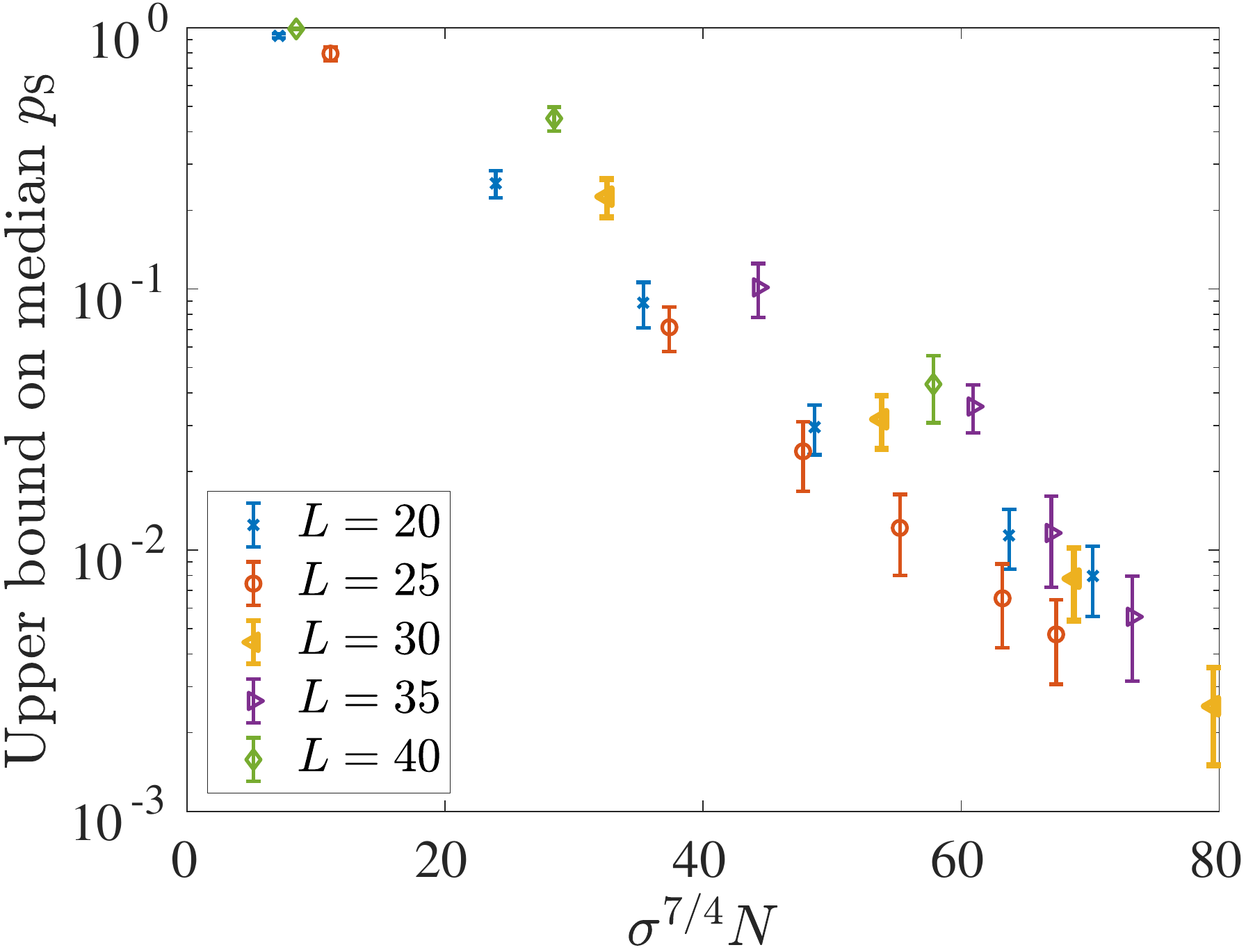} }
   \subfigure[]{\includegraphics[width=0.32\columnwidth]{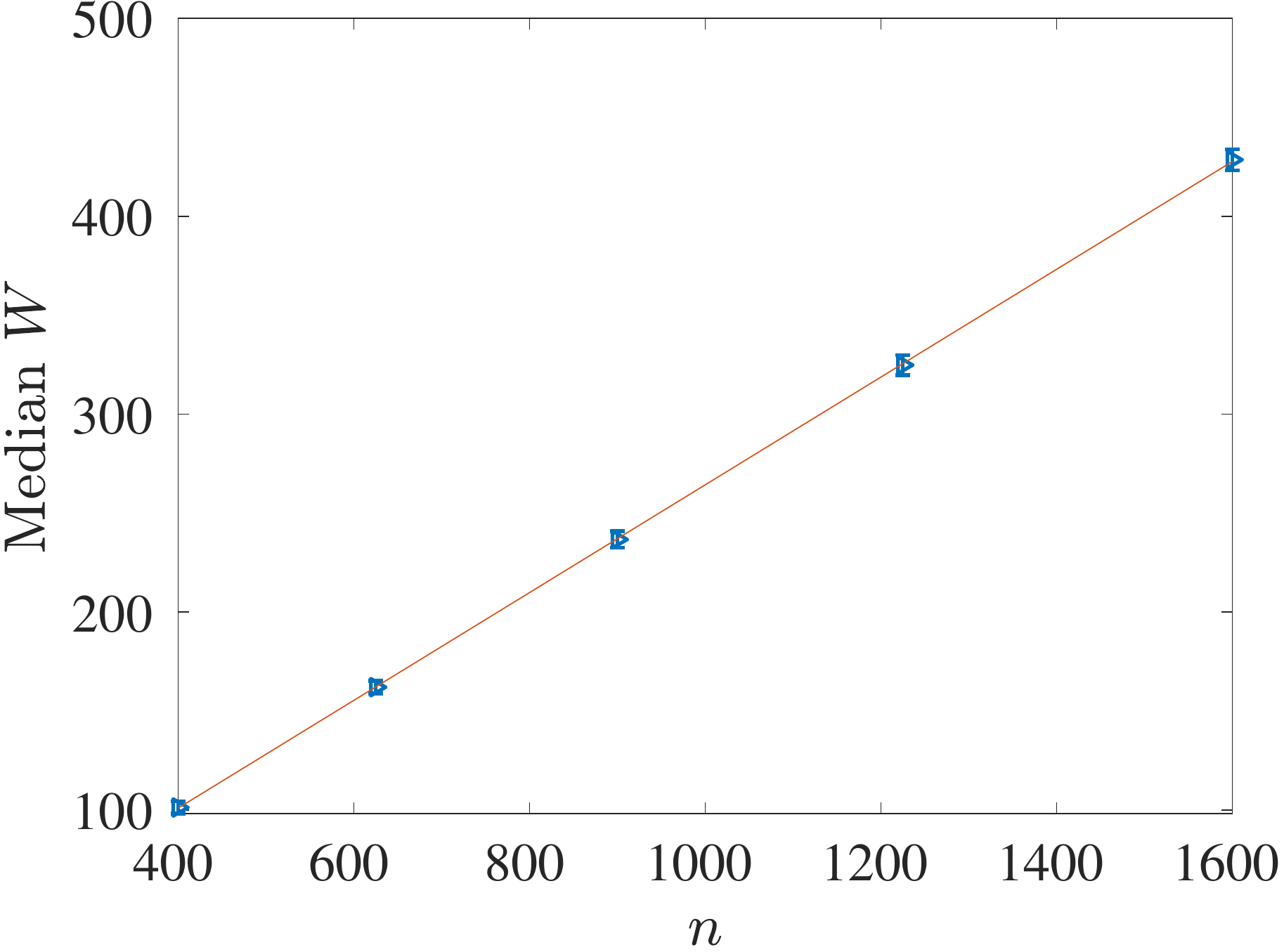}}
      \caption{(a) and (b) Upperbound on the median probability $p_{\mathrm{S}}$ that the ground state of the intended Hamiltonian remains the ground state of the implemented Hamiltonian for varying planted-solution instances on Ising $\pm 1$ instances defined on a square grid of linear dimension $L \in \left[ 20, 40 \right]$ and Gaussian noise strengths $\sigma \in \left[ 0.1,0.37\right]$.  For each instance, $p_{\mathrm{S}}$ is determined by averaging over 1000 noise realizations, and the median is taken over 100 instances. In (a), we use a scaling of $\sigma^2 n$, while in (b) we use a scaling of $\sigma^{7/4} n$.  (c) Median mass $W$ for topologically non-trivial ESs that are more optimal than the GSs for $\sigma = 0.2$. Linear fit given by: $W =  -7.84+ 0.27 n$.  Error bars correspond to two standard deviate error bars obtained from bootstrapping over the 100 instances.}
   \label{fig:SquareGrid_W_D}
\end{figure*}

We show in Figs.~\ref{fig:FittingToDataSquare} and \ref{fig:SquareGrid_W_D} our results.  The median probability $p_{\mathrm{S}}$ scaling is not as sharp as our previous examples, so we show two scalings $\sigma^2 n$ and $\sigma^{7/4}n$, with the latter becoming a better fit at smaller probabilities as shown in Fig.~\ref{fig:FittingToDataSquare}.  The variation in the data might be because these instances have many more GSs and show more variation in the absence of local fields.  Also, because the algorithm only finds one of the intended GSs, our value of $W$ may not be the smallest value, meaning that there may be other GSs for which $W$ is smaller.  

\section{Linear chain} \label{app:Chain}
%
We give here a simple derivation for the uniform chain of length $n$ with
coupling strength $J$.  We consider for concreteness the anti-ferromagnetic
case, but the argument equally applies to the ferromagnetic case. The GS
consist of an alternating chain ${+-+-+-\ldots}$ while the low-energy
excitations consist of solutions with domain walls. 
Let us consider a noise model where independent Gaussian noise $\delta J_{i, i+1} \sim \mathcal{N}(0, \sigma^2)$ occurs on each coupler:
\beq
J \to J + \delta J_{i,i+1} \ .
\eeq
There are $n-1$ such terms. The implemented ground state is different from the intended ground state when any of the couplings changes sign, i.e. when at least one coupling satisfies $J~+~\delta J_{i,i+1} ~<~0$.  Let us denote the probability of a single coupler changing sign by $p(z)$, where $z = J/\sigma$, which is given by:
\beq
p(z) = \frac{1}{\sqrt{2 \pi}} \int_{- \infty}^{-z} \mathrm{d} \eta e^{-\eta^2/2} \ .
\eeq
Therefore, the success probability, i.e. the probability that the implemented ground state is the same as the intended ground, is given by:
\beq
p_{\mathrm{S}}(z) = \left( 1 - p(z) \right)^{n-1} \ .
\eeq
Since $p(z) < 1$ for any finite $J$ and $\sigma$ values, the success
probability decreases exponentially with the chain length. However, this is
\emph{not} disorder chaos, because the noisy Ground State differs from the
clean one only \emph{locally} (i.e. the link-overlap between the two
Ground States is of order one).  These would correspond to what we referred to in the main text as topologically trivial states, which can be easily corrected to reach the clean Ground State.

Indeed, the problem of how much one can disorder an (anti)ferromagnetic system
before complex behavior appears has been
long-studied~\cite{imry:75,nishimori:80}. Generally speaking, the
ferromagnetic phase is rather resilient against disorder, which means that
complex Glassy behavior appears only when $\sigma\sim J$. For smaller
$\sigma$ the dirty Ground State do not differ in a significant way from the
clean Ground State, which probably explains why systems in a
(anti)ferromagnetic phase do not pose difficult optimization problems. In
fact, difficult optimization problems are expected to be in the glassy phase,
where ``disorder chaos'' is the rule.

\onecolumngrid


\end{document}